\documentclass[sigconf]{acmart}

\usepackage{balance}
\usepackage{longtable}
\usepackage{listings}
\usepackage{xcolor}
\AtBeginDocument{%
  \providecommand\BibTeX{{%
    \normalfont B\kern-0.5em{\scshape i\kern-0.25em b}\kern-0.8em\TeX}}}

\copyrightyear{2023}
\acmYear{2023}
\setcopyright{acmlicensed}\acmConference[CHI '23]{Proceedings of the 2023 CHI Conference on Human Factors in Computing Systems}{April 23--28, 2023}{Hamburg, Germany}
\acmBooktitle{Proceedings of the 2023 CHI Conference on Human Factors in Computing Systems (CHI '23), April 23--28, 2023, Hamburg, Germany}
\acmPrice{15.00}
\acmDOI{10.1145/3544548.3580919}
\acmISBN{978-1-4503-9421-5/23/04}

\begin{document}

\title{Studying the effect of AI Code Generators on Supporting Novice Learners in Introductory Programming}

\author{Majeed Kazemitabaar}
\orcid{0000-0001-6118-7938}
\affiliation{%
  \institution{Department of Computer Science, University of Toronto}
  \city{Toronto}
  \state{Ontario}
  \country{Canada}
}
\email{majeed@dgp.toronto.edu}

\author{Justin Chow}
\orcid{0000-0001-9668-8759}
\affiliation{%
  \institution{Department of Computer Science, University of Toronto}
  \city{Toronto}
  \state{Ontario}
  \country{Canada}
}
\email{justinchow@dgp.toronto.edu}

\author{Carl Ka To Ma}
\orcid{0000-0002-5178-6918}
\affiliation{%
  \institution{Department of Computer Science, University of Toronto}
  \city{Toronto}
  \state{Ontario}
  \country{Canada}
}
\email{cma@dgp.toronto.edu}

\author{Barbara J. Ericson}
\orcid{0000-0001-6881-8341}
\affiliation{%
  \institution{School of Information, University of Michigan}
  \city{Ann Arbor}
  \state{Michigan}
  \country{USA}
}
\email{barbarer@umich.edu}

\author{David Weintrop}
\orcid{0000-0002-3009-3899}
\affiliation{%
  \institution{College of Education, University of Maryland}
  \city{College Park}
  \state{Maryland}
  \country{USA}
}
\email{weintrop@umd.edu}

\author{Tovi Grossman}
\orcid{0000-0002-0494-5373}
\affiliation{%
  \institution{Department of Computer Science, University of Toronto}
  \city{Toronto}
  \state{Ontario}
  \country{Canada}
}
\email{tovi@dgp.toronto.edu}

\renewcommand{\shortauthors}{Majeed Kazemitabaar et al.}

\begin{abstract}
AI code generators like OpenAI Codex have the potential to assist novice programmers by generating code from natural language descriptions, however, over-reliance might negatively impact learning and retention. To explore the implications that AI code generators have on introductory programming, we conducted a controlled experiment with 69 novices (ages 10-17). Learners worked on 45 Python code-authoring tasks, for which half of the learners had access to Codex, each followed by a code-modification task. Our results show that using Codex significantly increased code-authoring performance (1.15x increased completion rate and 1.8x higher scores) while not decreasing performance on manual code-modification tasks. Additionally, learners with access to Codex during the training phase performed slightly better on the evaluation post-tests conducted one week later, although this difference did not reach statistical significance. Of interest, learners with higher Scratch pre-test scores performed significantly better on retention post-tests, if they had prior access to Codex.
\end{abstract}




\begin{teaserfigure}
  \includegraphics[width=\textwidth]{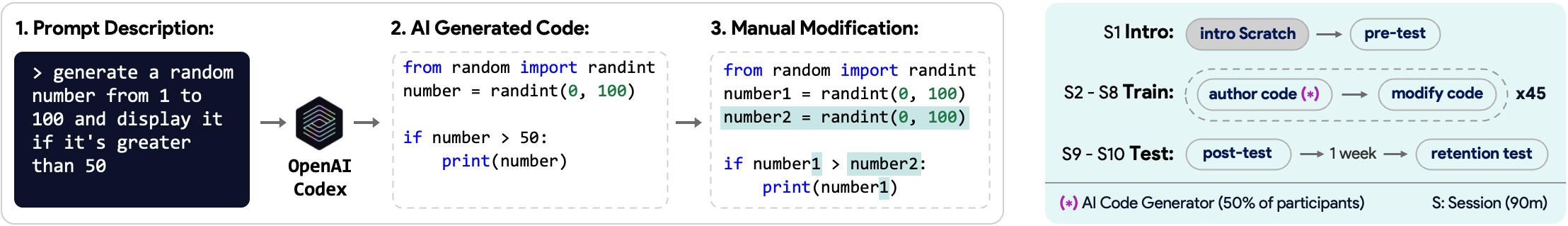}
  \caption{Left) Generate-modify usages with AI code generators. Right) Summary of our controlled study over 10 sessions.}
  \Description{The figure is divided into two parts. On the left, there is a description of a computer program that is sent to OpenAI Codex, which generates code. The AI-generated code is displayed, which checks if a random number between 0 and 100 is greater than 50. The user modifies the code by adding another random number and checking if the first number is greater than the second. This process is a common generate-modify usage with AI code generators. On the right, the figure shows a summary of a controlled study conducted over 10 sessions. The first row summarizes the introductory Scratch session and the pre-test evaluation. The second row includes 45 two-part programming tasks over 7 sessions. The last row shows 2 evaluation tests: an immediate post-test and a retention post-test.}
  \label{fig:teaser}
\end{teaserfigure}

\maketitle

\section{Introduction}
Powered by the recent advancements in Deep Learning \cite{vaswani2017attention}, Large Language Models that are trained on millions of lines of code, such as OpenAI Codex \cite{chen2021evaluating}, can generate code from natural language descriptions (Figure~\ref{fig:teaser}, Left). In addition to enabling natural language programming, these AI coding assistants can perform numerous operations including code-to-code operations like code completion, translation, repair, and summarization, along with language-to-code operations such as code explanation and search \cite{lu2021codexglue,sarsa2022automatic}. By generating code from simple sentences instead of formal and syntactically fixed specifications, these AI Coding Assistants may lower the barriers to entry into programming.

In the context of computer science education, AI code-generators could potentially support both learners and educators. For instance, code generators could automatically fix semantic bugs and syntax errors, and allow learners to focus more on theoretical and problem-solving aspects of computational thinking \cite{wing2006computational}, instead of struggling with syntax. Additionally, these tools could support educators in curriculum development by generating programming exercises and explanations to solutions \cite{sarsa2022automatic}.

However, there are also several potential drawbacks in using such tools in instructional contexts: learners might become dependent on these tools and not be able to author similar code without them; they might not know how to best express their intentions in order to generate their desired code \cite{vaithilingam2022expectation} and/or, they might not understand the AI-generated code and how they could modify it if needed. From the educator’s point of view, there are also concerns related to academic integrity and plagiarism \cite{finnie2022robots}.

Prior work has explored AI code-generators from a usability perspective with experienced programmers \cite{vaithilingam2022expectation}, usage barriers for novice programmers, and advantages to curriculum development for educators \cite{sarsa2022automatic}. However, AI code generators have not been studied from a learning point of view in an introductory programming context, thus leaving essential and foundational questions unanswered. Central among them is understanding whether novices who have never written text-based code are able to understand the code generated by these tools, and if they are able to modify or extend the generated code. It is also important to understand if using such tools will form a reliance, or help learners write code without such tools being present. Specifically, we are interested in answering the following questions:
\begin{itemize}
    \item \textbf{RQ1:} Are novices able to utilize AI code generators to solve introductory programming tasks?
    \item \textbf{RQ2:} How do learners’ task performance measures (e.g., correctness score, completion time, and error rate) differ with and without AI code generators?
    \item \textbf{RQ3:} How does learners’ ability to manually modify code differ with and without the use of AI code generators?
    \item \textbf{RQ4:} What are the effects on learning performance and retention from using an AI code generator versus not?
    \item \textbf{RQ5:} What is the relationship between existing programming competency and the effectiveness of using AI code generators in introductory programming?
\end{itemize}

To answer these questions, we conducted a controlled study with 69 young learners, ages 10-17 (\textit{M}=12.5, \textit{SD}=1.8) with no prior text-based programming experience. Half of them used an AI Coding Assistant (Codex) to learn Python, and the other half did not. Specifically, we built Coding Steps, a self-paced learning environment that included novice-friendly documentation in which learners worked on small and increasingly complex programming tasks to learn the basic concepts of Python programming. Concepts including variables, operators, data-types, conditionals, loops, and arrays. During the three-week study, learners went through three phases (Figure~\ref{fig:teaser}, Right): (i) a single \textit{introduction} session to the basic concepts of programming using Scratch, followed by a pre-study evaluation, (ii) seven \textit{training} sessions on introductory-level Python programming topics by working on 45 code-authoring tasks, each followed by a code-modification task, and (iii) two \textit{evaluation} sessions, including an immediate post-test session, followed by a retention test conducted a week later.

Our results show that learners who had access to the AI code generator (the Codex group) were able to successfully generate code and showed evidence of understanding the generated code during the training phase. They performed significantly better on code-authoring tasks (1.15x increased progress, 0.59x less errors, 1.8x higher correctness, and 0.57x less time) with no decrease in performance on the following manual code-modification tasks, in which both groups performed similarly. Furthermore, during the evaluation phase and on the immediate post-test, learners from the Codex group were able to perform similar to the Baseline group despite not having access to the AI code generator. In the retention test which was conducted one week later, learners from the Codex group performed slightly better on coding tasks and multiple-choice questions, although these results did not reach statistical significance. Finally, our analysis indicates that learners with more prior programming competency may benefit more from AI code generators. After discussing these results, we conclude with a discussion of limitations of our work and the implications of our results.

\section{Related Work}
Our work draws from several areas of prior research: natural language programming, AI Coding Assistants powered by large language models, and introductory programming. We review each in turn.

\subsection{Natural Language Programming}
Historically, many computer scientists have been arguing that programming a computer should remain a completely formal process with syntactically fixed specifications, similar to math, as it would encourage people to think more like \textit{computers} and therefore, perform better \cite{dijkstra1979foolishness}. In contrast, there are also arguments supporting the idea of bringing programming closer to how \textit{people} think \cite{myers2004natural,pane2006more} through natural language programming. From an HCI point of view, expressing algorithms, data manipulation tasks, and even debugging, require many transformations from how people think to how such concepts are specified in the language of the machine \cite{pane2002programming}. By reducing this gap, and utilizing principles like direct manipulation \cite{hutchins1985direct}, the transformation effort could potentially be reduced and lighten up the burden of programming.

Generating code from natural language, instead of syntactically fixed and formal specifications, comes with many challenges like handling ambiguity and abstract language. Various methods have been employed, such as using semantic parsers \cite{ballard1979programming,begel2005spoken,biermann1983experimental,desai2016program,gulwani2011automating,gulwani2014nlyze,knoll2006pegasus,landhausser2017nlci,le2013smartsynth,price2000naturaljava,little2006translating,schlegel2019vajra} and machine learning \cite{balog2016deepcoder,ling2016latent,quirk2015language,raghothaman2016swim,raza2015compositional,yin2017syntactic,zhong2017seq2sql}. However, initial approaches were mostly designed to handle line-by-line explanations instead of abstract explanations.

\subsection{AI Coding Assistants}
Recently, there have been significant advancements in Deep Learning and Large Language Models (LLM) like GPT-3 \cite{brown2020language}. These models have been able to generate code from natural language descriptions when they are trained on large corpora of source code. For example, models like OpenAI Codex \cite{chen2021evaluating}, Microsoft CodeBERT \cite{feng2020codebert}, Google PaLM \cite{chowdhery2022palm}, DeepMind AlphaCode \cite{li2022competition} have been trained on millions of lines of publicly available code (e.g., on Github). In addition to generating code from descriptions, these tools can perform code completion, translation, repair, summarization, and explanation \cite{lu2021codexglue}. Currently, these models power AI Coding Assistants including Github Copilot \cite{copilot}, CodeWhisperer \cite{codewhisperer} Tabnine \cite{tabnine} that provide code-completion functionality.

To explore the explainability needs of AI code generators, Jiao et al. \cite{sun2022investigating} conducted 9 semi-structured workshops with 43 software engineers using AI Coding Assistants and identified 11 categories of explainability needs such as types of input that the model can take, and the data that these models are trained on. In a study with 24 experienced programmers that used Github Copilot to complete real-world programming tasks, Vaithilingam et al. \cite{vaithilingam2022expectation} reported that using Copilot did not improve task completion time or success rate, however, most programmers preferred to use it in their daily programming tasks as it provided a useful starting point. Jiang et al. created GenLine \cite{jiang2022discovering} and conducted a study with 12 participants that worked on two web-programming tasks to explore the user experience of natural language programming with AI code generators. Their results indicate that participants generally felt that they needed to learn the syntax of natural language programming. Finally, in a recent study to explore the learnability of program synthesizers, Jayagopal et al. \cite{jayagopal2022exploring} found that participants preferred program synthesis tools where users can also write code manually and liked the \textit{triggerless initiation} and \textit{triggerless result communication} mechanisms of Copilot. Unlike these previous studies, we are particularly interested in the impact that AI Coding Assistants have on novices when they are first learning text-based programming languages and conduct the first study comparing learning experiences with and without an AI code generator in a controlled study.

\subsection{Introductory Programming}
Computing is now being integrated into K-12 education of many countries \cite{popat2019learning, webb2017computer}. Research has provided evidence that computing education offers a platform to practice and improve problem-solving \cite{fessakis2013problem,kalelioglu2014effects,miller1988effects,psycharis2017effects}, critical thinking \cite{falloon2016analysis}, collaboration \cite{resnick2015different}, and active learning \cite{kaleliouglu2015new} skills. One of the core components of CT is \textit{programming}, which supports the cognitive tasks involved in CT and demonstrates computational competency \cite{grover2013computational}. A full review of introductory programming research is beyond the scope of this paper, and instead we refer the reader to recent surveys on the topic \cite{becker201950,luxton2018introductory}. Here we review introductory programming research that is most relevant to our own work: learning challenges, assistive programming environments, and AI code generators.

Programming is not easy to learn \cite{du1986some,qian2017students}. Novices can be overwhelmed by the complexity of coding tasks \cite{kinnunen2010experiencing,van2003taking} and spend an unexpectedly large amount of time on them \cite{benda2012life}. This can be a frustrating experience \cite{rodrigo2009coarse} for students and repeated failures, especially early on, can lower their self-efficacy with respect to programming \cite{kinnunen2011cs}. Cognitive load theory defines the demand that a situation or task places on a learner’s working memory \cite{sweller2019cognitive} and is based in part on the learner’s prior knowledge and the complexity of the task or material \cite{duran2018towards}. Several approaches have been developed to reduce this cognitive load in introductory programming, such as the use of worked examples and similar practice problems \cite{renkl2005worked,van2003taking}, or mixed-up code (Parsons) problems \cite{parsons2006parson}, where learners are given mixed up fragments that must be placed in the correct order to solve a problem. Learner’s typically complete Parsons problems significantly more quickly than fixing or writing code and have similar learning gains \cite{ericson2018evaluating,ericson2017solving,zhi2019evaluating}. Similarly, AI code generators may reduce the cognitive load when learning to program, but its impact on the learning experience has not been studied.

One main approach to improving the learning experience is the use of assistive programming environments, which can help alleviate misconceptions in syntax \cite{altadmri201537,hristova2003identifying}, and conceptual knowledge \cite{sirkia2012exploring}. For example, many introductory programming courses for K-12 learners start with block-based programming environments (BBPEs) like LogoBlocks \cite{begel1996logoblocks}, Alice \cite{cooper2000alice}, Scratch \cite{resnick2009scratch}, and AppInventor \cite{wolber2011app}. BBPEs have been designed to eliminate syntax errors and enable students to work on personally meaningful projects \cite{resnick2014give}. These characteristics lower the barrier of entry to programming and allow learners to focus on learning how to formulate a solution that a machine can execute. However, learners may perceive BBPEs to be less powerful and wish (or need) to transition to text-based programming languages. This transition, however, comes with its own challenges. To support this transition, various tools have been developed, including: dual-modality programming environments like PencilCode \cite{bau2015pencil} and MakeCode \cite{ball2019microsoft}, and hybrid environments such as Frame-based editing \cite{kolling2015frame} and CodeStruct \cite{kazemitabaar2022codestruct,kazemitabaar2023scaffolding}. In our study, we examine if AI code generators may have similar learning benefits to BBPEs and evaluate a subsequent transition to manual text programming. 

With the increased availability of AI code generators like OpenAI Codex and Github Copilot, researchers have started to explore the implications of such tools on introductory programming. For example, Finnie-Ansley et al. \cite{finnie2022robots} showed that OpenAI Codex outperforms most students on code writing questions that are used in typical first-year programming exams. From an educator’s point of view, Sarsa et al. \cite{sarsa2022automatic} qualitatively analyzed the novelty, sensibleness, and readiness of 120 programming exercises that were generated by OpenAI Codex after priming it with a few samples. Their results showed that the majority of the programming exercises created by Codex were sensible, but generated exercises needed some adjustments. From the learner’s perspective, there are still many open questions: What happens if learners have access to AI Coding Assistants during their training? Are novices able to use these tools and understand the generated code? Will they become over-reliant on AI-generated code? We seek an answer these, and other related research questions, in this paper.

\begin{figure*}
    \centering
    \includegraphics[width=\textwidth]{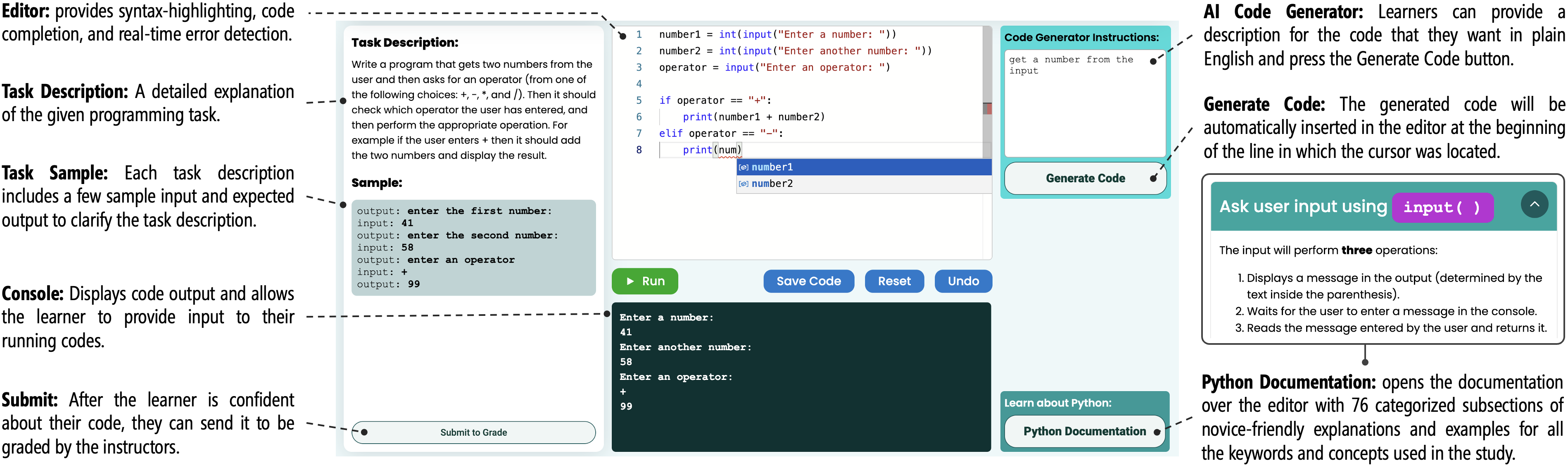}
    \caption{Coding Steps Programming environment, AI code generator, and python documentation.}
    \Description{The figure illustrates the interface of the Coding Steps programming environment, consisting of multiple components. On the left of the interface, there is a task description and a set of examples. To complete the task, users must work on the code and submit it. The middle of the interface contains a code editor that features syntax highlighting and real-time error detection. Beneath the code editor is a console, which displays output and allows for user input to the running code. In the top-right corner of the interface, there is an input textbox that enables the user to write natural language descriptions of behaviors and request the tool to generate code using OpenAI Codex. At the top of the figure, there is a documentation button that opens a Python documentation on top of the entire interface as an overlay when clicked.}
    \label{fig:coding_steps_system}
\end{figure*}

\section{AI-Assisted Learning Environment}
To investigate the effect of using AI Coding Assistants when learning to code, we built Coding Steps, a web-based application that enables learning of basic Python programming. The system allows learners to work on a series of programming tasks that were designed to gradually introduce new concepts. The system includes functionality to submit code to remote instructors that grade submitted work and provide feedback through a grading dashboard. Figure~\ref{fig:coding_steps_system} shows the programming environment in Coding Steps that includes: (i) a description of the task and a set of examples, (ii) a code editor with syntax highlighting, real-time error detection, and autocomplete functionality, (iii) a console to display the output and provide input to the running code, (iv) an embedded Python documentation with mini examples specifically written for novices, and (v) a code generator for inserting AI-generated code into the code editor. The integration of all these features in one single application allows learners to independently progress through the training tasks, while also supporting the collection of usage data for our study. The source code of Coding Steps is available as an open-source repository including both client and server-side code: \href{https://github.com/MajeedKazemi/coding-steps}{https://github.com/MajeedKazemi/coding-steps}.

\subsection{Implementation}
Coding Steps is written in TypeScript and has a client-server architecture that enables authentication and storing user progress, collecting logs, providing personalized feedback, executing code, and communicating with OpenAI Codex, the AI code generator used for our study. The server was implemented using Node.js, specifically: Express.js for REST API used in client-server communication, Mongoose to interact with a cloud-based instance of MongoDB for storing user progress, Passport.js for user authentication, python-shell for running Python code on the server, and a custom Python language server for real-time autocomplete suggestion and error detection. The client-side code was developed using the React Framework And included the Monaco Editor for editing code and syntax-highlighting. The Monaco Editor communicated with the Python Language Server to provide code completion, real-time error detection, definitions, hover, and references inside the editor. The Python documentation was designed as a pop-up window that would cover most of the interface. Its content was broken down into multiple subsections, each covering a specific concept of Python programming. This allowed for detailed collection of documentation usage metrics during the study.

\begin{table*}[t]
\centering
\caption{Tasks used in Coding Steps broken down by topic}
\label{tab:tasks_by_topic}
\begin{tabular}{l l l l}
    \toprule
    \textbf{Topic} & \textbf{\# Coding Tasks} & \textbf{\# MCQs} & \textbf{Python Concepts}\\
    \midrule
    Basics & 8 & 6 &input / output, variables, basic operators, joining strings, random numbers\\
    Data-Types & 4 & 4 & type conversions (from string to integer and vice versa)\\
    Conditionals & 8 & 10 & conditional statements, logical expressions, comparators\\
    Loops & 18 & 9 & iterators, loops, and conditional loops\\
    Arrays & 7 & 10 & indexing lists, appending items to lists, obtaining the length of lists\\
  \bottomrule
\end{tabular}
\end{table*}

\begin{figure*}
    \centering
    \includegraphics[width=\textwidth]{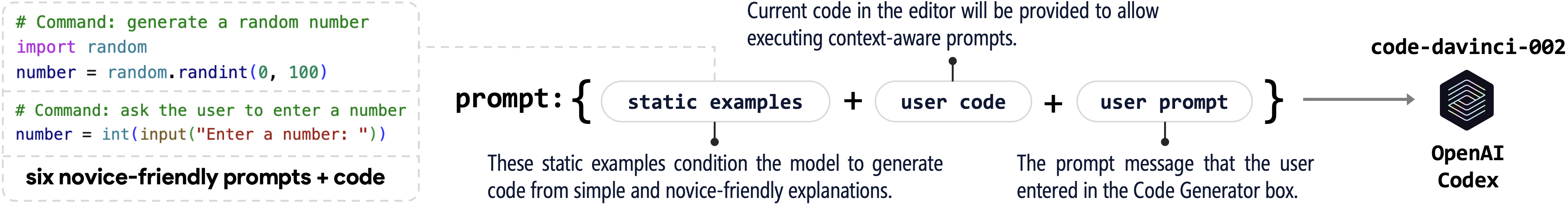}
    \caption{Prompts sent to the OpenAI Codex API were modified to additionally include current user code and six samples.}
    \Description{The figure depicts the process by which prompts are sent to the OpenAI Codex model for code generation. The prompt consists of three main parts, which are used as a few-shot learning sample. The first part is six static examples, which are designed to condition the model to generate code from simple and novice-friendly explanations. The second part is the user's code, which allows for context-aware code generation. Finally, the third part is the user's prompt. All three parts are aggregated and sent to the OpenAI Codex API, which generates code based on the user's prompt.}
    \label{fig:prompt_construction}
\end{figure*}

For AI code generation, we included a textbox and Generate Code button next to the code editor (Figure~\ref{fig:coding_steps_system}, top right). Users could type their desired code behavior into the textbox using natural language. After clicking on the generate button, the code generated from OpenAI Codex would be inserted at the beginning of the line in which the cursor was located (and shifting any existing code below). The code generator uses the \textbf{code-davinci-002} model from OpenAI’s code completion API that generates code from the provided prompt. The prompt message was modified on each API call to include the following three parts to help seed the code generation process (Figure~\ref{fig:prompt_construction}): (i) six static examples of short prompt messages followed by desired output code (ii) the current code in the editor, and finally (iii) the user’s requested behavior. This would condition the AI model on generating python code from simple and novice-level explanations and allows the code generator to be aware of the user’s context. Coding Steps was approved by the OpenAI App Review team prior to running the study.

\subsection{Data Instrumentation}\label{sec:data_instrumentation}
The system was instrumented to collect timestamped usage data. From the code editor, all edits (including insertions, deletions, or replacements) were collected. From the console, all standard inputs, outputs, and errors were collected. From the Python documentation, all open and close events were collected from each of the subsections in the documentation. Finally, from the AI code generator, all prompt messages and generated code were collected. Data logs were periodically sent to a remote server (every 30 seconds), or upon submitting a task, to be stored in a database.

\subsection{Programming Task Design}
The learning environment included 45 two-part programming tasks that were designed to avoid overwhelming learners through too much cognitive load \cite{duran2022cognitive} by gradually increasing complexity and introducing new concepts as they made progress. This was intended to keep learners in the Zone of Proximal Development \cite{vygotsky1978mind}, which means learners were challenged to do more than they could without help. Learners’ Zone of Proximal Development expands as they learn, and therefore, they can work on more complicated tasks after they master easier tasks. The tasks were organized into five groups that each focused on one specific topic in Python. Each group of programming tasks was followed by several multiple-choice questions (MCQs) for additional practice (though the correct answer was never given). See Table~\ref{tab:tasks_by_topic} for more information about the tasks and Appendix A for further detail on each of the task descriptions.

\subsection{Quality of AI Generated Code}
To measure the quality of the AI code generator, we entered the task description of the 45 code-authoring tasks into the pre-conditioned model (described in Figure~\ref{fig:prompt_construction}) and evaluated the accuracy and amount of required manual modifications on the generated code. Codex was able to correctly solve 41/45 of the tasks with no changes to the task description. For three tasks, it required minor modifications and rewordings to the prompt message, and for one task it did not import the random module when randint was used in the generated code (however for 11/12 of the other tasks that the random module was used, Codex did correctly import the random module). This shows that the AI code generator produces high quality results, but this depends on the quality of the prompt message.

\section{User Study}
We now present our study to investigate the impact of using AI Coding Assistants when learning to code. The study was specifically focused on novice programmers and learning introductory Python programming skills. Students used Coding Steps to learn independently throughout the study, receiving assistance from the experimenters (first three authors) when needed. Half of the students had access to the AI code generator during the training phases of the study.

\subsection{Study Procedure}
\begin{figure*}
    \centering
    \includegraphics[width=\textwidth]{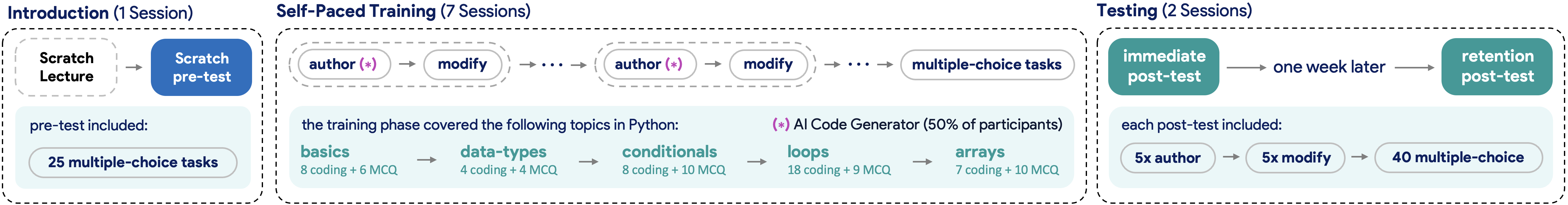}
    \caption{Study procedure over the 10 sessions (three weeks).}
    \Description{The figure is a detailed visual representation of the study procedure that spans over 10 sessions (in three weeks). The procedure comprises three main phases. The first phase is the introduction section that includes a Scratch lecture and a Scratch pre-test. The pre-test consists of 25 multiple-choice questions and is completed in one session. The second phase is the training phase, which spans over seven sessions. In this phase, participants were assigned 45 two-part programming tasks. Half of the participants had access to an AI code generator. The third and final phase is the evaluation phase, which includes an immediate post-test completed in one session and a retention test completed after one week in another session. Each post-test consists of 5 code-authoring tasks, 5 code-modification tasks, and 40 multiple-choice questions.}
    \label{fig:study_procedure}
\end{figure*}

A between-subject design was used, with two conditions: one in which students had access to the AI code generator (Codex) and one in which the code generator was removed (Baseline). The study lasted three weeks and included ten 90-minute sessions, with one session per day, spanning over three (consecutive) weeks. The study was broken into three main phases (Figure~\ref{fig:study_procedure}): an introduction phase, a training phase, and an evaluation phase. The study was conducted remotely over the Google Meet platform.

\subsubsection{Introduction Phase}
The first session in the training phase was 2-hours long and included a one-hour introduction to the basic concepts of programming and computational thinking using Scratch. The lecture covered the following topics: sequence, input and output, random numbers, joining texts, arithmetic operations, conditionals, logical expressions, loops, conditional loops, and lists. During the second hour of the first session all participants worked on a pre-test including 25 multiple-choice questions about Scratch programming (Figure~\ref{fig:task_samples}a). 

\subsubsection{Training Phase}
The training phase consisted of seven sessions of using Coding Steps. Learners from both conditions worked on 45 programming tasks and 40 multiple-choice questions. Students progressed at their own pace through five main topics during these sessions: basics (including variables, input/output, operators, and random numbers), data-types, conditionals, loops, and lists. Learners were encouraged to ask questions from the instructors if they needed help and the instructors were trained to guide the learners to the specific parts of the documentation to resolve their issues. Additionally, learners received personalized feedback for each of their incorrect submissions that were rejected. This feedback was displayed inside the Coding Steps environment for easy access.

\begin{figure*}
    \centering
    \includegraphics[width=\textwidth]{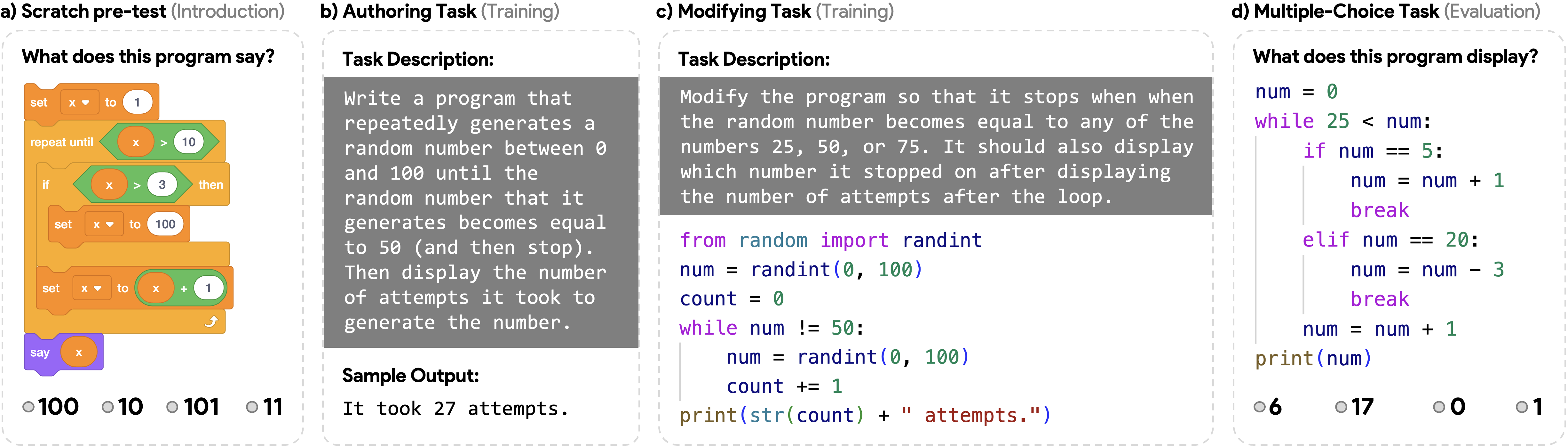}
    \caption{Sample tasks used in the pre-test, training, and evaluation phases.}
    \Description{The figure depicts four different types of tasks that were utilized in various stages of our study. The tasks are as follows: (1) A multiple-choice question that involves scratch programming. The question features an "if" block nested inside a "repeat until" block, and there are four possible choices presented below the code. (2) A code authoring task, which is accompanied by a task description and a sample output. (3) A code-modification task, which includes a task description and the code that needs to be modified. (4) A Python multiple-choice question that asks what a particular code would display.}
    \label{fig:task_samples}
\end{figure*}

Learners began the training phase by watching a video that explained how the Coding Steps system worked. The Codex group watched a different video that included several examples of AI code generation. After watching the video, learners could start working on the tasks. Each programming task in the learning phase had two parts: authoring code (Figure~\ref{fig:task_samples}b) and modifying code (Figure~\ref{fig:task_samples}c). For both types of tasks, learners first read the task description and a few examples of input and expected output. They were then asked to write code in the code editor and were free to use the Python documentation when needed. Additionally, learners in the Codex condition had access to the AI code generator. The tasks and the topics were presented in a fixed order for all students in both groups, as they were designed to gradually increase in complexity.

\begin{figure*}
    \centering
    \includegraphics[width=\textwidth]{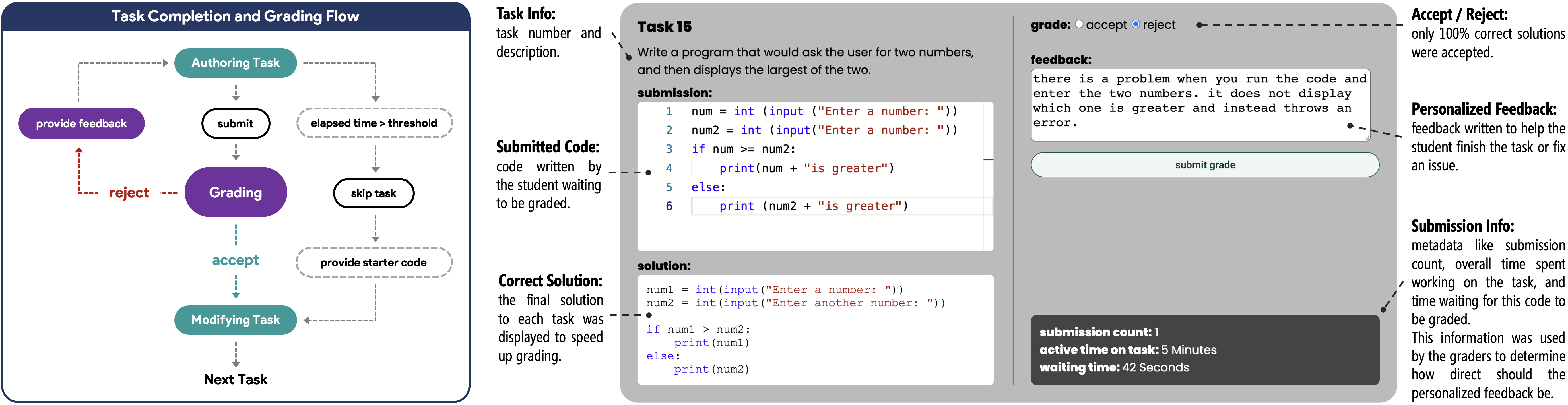}
    \caption{Left) Task completion and grading flow during the training phase with Coding Steps. Right) The grading UI that was used in the grading dashboard which includes anonymous information for each code submission.}
    \Description{The left part of this figure shows a state diagram that illustrates how tasks are graded in our study. When an authoring task is submitted, graders will either accept or reject it. If the task is accepted, the modifying task will start with the user's own solution. However, if it gets rejected, the instructors will provide feedback, and the diagram goes back to the "authoring task" stage, which means that the learner has to continue working on the task. If learners are stuck and the elapsed time is greater than a particular threshold, they can also skip the task. This would take them to the modify part of the task, however, the starter code will be provided for them in this scenario. On the right side of this figure, the grading UI used in the admin dashboard of Coding Steps is displayed. The interface shows the submitted code, the task description, and the correct solution. It does not show any identifiable information, such as the name or group condition, which helps both groups receive consistently similar personalized feedback.}
    \label{fig:task_flow}
\end{figure*}

 The code submission and grading process is illustrated in Figure~\ref{fig:task_flow}, Left. After running the code to check that it worked, learners submitted the code in order to advance to the next task. At this point, the researchers conducting the study would synchronously, either accept a correct solution or send an incorrect or incomplete solution back to the learner along with personalized feedback. Learners were required to continue working on the task before advancing to the next task. However, learners were also able to skip tasks after exceeding the minimum amount of time (which varied between 3 to 12 minutes based on task difficulty). If the learner skipped an authoring task, they would be shown the correct solution which was then used as the starter code for the associated code modification task. If their submission was correct, their own accepted submission was used as the starter code for the code modification task. To ensure both groups received consistent and unbiased feedback on their submissions, the grading dashboard did not display any information that would reveal the identity of learners or their group (Figure~\ref{fig:task_flow}, Right). A first-in, first-out mechanism was used to ensure submissions were graded in order of submission. For each submitted task that was incorrect, graders were instructed to initially just explain what was wrong with the submission. However, as the number of re-submissions or the amount of time that a learner had spent working on a task increased, graders were instructed to provide more direct feedback that included some hints.

\subsubsection{Evaluation Phase}
The immediate post-test was conducted one day after the training phase. The Coding Steps application was still used for the evaluation phase, but learners did not have access to either the code generator or the Python documentation, or any feedback after submitting the tasks. The tasks in this phase included five code-authoring tasks, five code modification tasks, and then 40 multiple-choice questions (Figure~\ref{fig:task_samples}d). In the code-authoring tasks, learners had to manually write all code, while in the modification tasks, starter code was provided, and learners were asked to change it. Generally, the evaluation tasks were analogous to the tasks in the training phase in terms of difficulty and topics. The second evaluation session, the retention test, was conducted one week later and included a new set of tasks and questions with a similar number and order of tasks to the immediate test.

\subsection{Participants}
Participants were recruited through multiple coding camps located in two major North American cities. From more than 200 sign-ups, we contacted 90 learners that reported no prior text-based programming experience to participate in the study. From the 90 participants that started the study, 69 learners (21 female/48 male) ages 10-17 (\textit{M}=12.5; \textit{SD}=1.8) successfully completed all three phases of the study (11 only participated in the first session, four participated in less than half of the sessions, and six missed one of the last two evaluation sessions). No common factors were identified among the 21 students who dropped out in terms of disability, native language (six non-English) and computer or internet access. Only the results from the 69 learners that completed the study are included. Participants received a \$50 gift card as compensation and the study protocol was approved by our institution’s Research Ethics Board. To accommodate all learners with different time constraints, three sections were offered each day of the study and learners could choose which one to join each day. The first nine sessions were conducted every day across two calendar weeks (there was no session on weekends), with the final retention post-test conducted the following week.

None of the participants reported any prior text-based programming experience, 64 indicated using a block-based programming environment like Scratch or Code.org, and 27 indicated taking a programming-related class in the past. English was the native language for 51 participants and five reported that explaining things in English was difficult for them (four of them were native English speakers). Three participants each indicated having one disability: ADHD, vision impairment, and hearing impairment. Although socioeconomic status was not asked about directly, all participants had access to a personal computer (51 indicated that they had their own computer) and 41 had at least two personal devices (e.g., a computer and a tablet or a phone).

\subsubsection{Participant Condition Assignment}
Following the Introduction Phase using Scratch, participants were divided into two groups using a matched-groups design \cite{bruhn2009pursuit,mackenzie2012human}. The ninety participants who completed the first session were divided into two groups to have similar means and variances based on their pre-test scores on Scratch. This process was used to balance the Codex and Baseline groups with regard to prior programming knowledge. One learner in each pair was randomly assigned to either the Codex or Baseline group.

Accounting for the 21 participants who did not complete the entire study, we ended up with 33 learners in the Codex group, and 36 learners in the Baseline group that finished all 10 sessions. These participants had similar means on prior Scratch programming knowledge as measured by the pre-test (Codex: \textit{M}=62.7\%; Baseline: \textit{M}=60\%; t(67)=0.54,  \textit{p}=.67).

\subsection{Data Collection}
To analyze learners coding performance, Coding Steps performed low-level instrumentation automatically, as described in Section~\ref{sec:data_instrumentation}. Additionally, demographic information was collected on the first session after the Scratch evaluation. Finally, a post-study survey was administered at the end of the study that included short answers and Likert questions about learners’ perceptions about the Python documentation, learning gains, confidence, and several questions about the code generator exclusively in the Codex group.

\begin{table*}[t]
\centering
\caption{Definition, unit, source, and calculation method for each of the metrics.}
\label{tab:metrics_def}
\begin{tabular}{p{0.3\linewidth} p{0.66\linewidth}}
    \toprule
    \textbf{Overall Training Metrics} & \textbf{Definition and Source} \\
    \midrule

   Completion Rate (percentage) & \textit{Definition}: How far a learner progressed through the training phase regardless of correctness of tasks or skips: number of seen tasks divided by total tasks count.\\

   Personalized Feedback (count) & \textit{Definition}: Total number of personalized feedbacks a learner received during the training phase. \textit{Source}: code submission logs.\\

   Feedback length (characters) & \textit{Definition}: The length (number of characters) of the personalized feedback a learner received during the training phase. \textit{Source}: code submission logs.\\

Direct hints (count)	& \textit{Definition}: Total number of personalized feedbacks a learner received that included direct hints towards solving the problem. \textit{Source}: code submission logs.\\

    \toprule
    \textbf{Per-Task Performance} & \textbf{Definition and Source} \\
    \midrule

Coding Correctness Score (percentage) & \textit{Definition}: How correct was a learner's solution to a single task. \textit{Source}: Final submission in the submissions log that was graded independently by two researchers.\\

MCQ Correctness Score (percentage) &	\textit{Definition}: Whether a learner responded correctly to a multiple-choice question. \textit{Source}: submission logs.\\

Completion Time	(seconds)	& \textit{Definition}: Active time a learner spent working on a task (by removing inactivity gaps of longer than one minute). \textit{Source}: aggregated logs.\\

Documentation Referenced (count)	& \textit{Definition}: Whether a learner referenced the Python documentation for a task or not. \textit{Source}: documentation logs.\\

Encountered Errors (count)	&	\textit{Definition}: Number of errors a learner encountered after running their code categorized into syntax, data-type, and semantic errors. \textit{Source}: console logs.\\

    \toprule
    \textbf{AI Code Generator Usage} & \textbf{Definition and Source} \\
    \midrule

Code Generator Usage Per Task	(count)	& \textit{Definition}: Number of unique prompts and codes generated using the AI code generator during a single task. \textit{Source}: code generator logs.\\

AI-Generated Code Ratio (percentage)	& \textit{Definition}: The percentage of code in a task that was generated by an AI code generator, as opposed to being written manually by the learner calculated using the Jaccard text similarity coefficient \cite{jaccard1901etude}. \textit{Source}: code submission logs and code generator logs.\\

Tasks Broke Down into Subgoals (count) & \textit{Definition}: Whether different parts of the final submission for a task was generated from different codex usages. \textit{Calculation Method}: averaging over the maximum hamming distance between each line of the final submission and each line in the codex generated codes. \textit{Source}: code submission logs and code generator logs.\\

Prompt Similarity with Task Description (percentage) &	\textit{Definition}: Similarity between the prompt used for generating code and the task description. \textit{Source}: code generator logs and task descriptions.\\

  \bottomrule
\end{tabular}
\end{table*}

\subsection{Data Analysis}
For calculating correctness scores on the coding tasks, two independent researchers graded each submitted solution using a simple and consistent grading rubric of deducting 25\% for each major issue or missing part in their final submission (0\%, 25\%, 50\%, 75\%, and 100\%). The two graders fully agreed on 79\% of the submissions, with 16\% of disagreements having a difference of 25\% in which we averaged the two grades. For the 5\% of the disagreements which were more than 25\%, the two graders met again to resolve each of their disagreements. A similar approach was used for checking if any of the personalized feedback provided during the training phase could be counted as a direct hint or not.

Here we define the metrics that we used to analyze learners’ performance and behavior while as they worked on the programming tasks. Our logging system enabled us to define various metrics that were all computed programmatically which can be seen in Table~\ref{tab:metrics_def}.

Overall Training Metrics: to measure overall task completion rate, we simply divided the number of tasks a learner completed or skipped, by the total number of tasks in the training phase as all learners went through a fixed order of tasks. To measure how much feedback learners received during the training phase, we quantitatively analyzed their length (in characters), count, and qualitatively analyzed whether they could be counted as a direct hint or not.

Task Performance Metrics: to measure task completion time, we calculated the active time a learner spent working on a task by removing inactivity gaps of longer than one minute. To check if a learner referenced the documentation during a task, we analyzed the open and close events on the subsections of the documentation. Furthermore, to categorize and count encountered errors during each task, the Python shell logs were scanned for three major types of errors: syntax, data-types, and semantic errors. Syntax errors consisted of issues related to indentation, mismatched identifiers, missing or mismatched quotes and parenthesis, incorrect or missing imports, missing or extra arguments in function calls, and general syntax errors. Data-type errors consisted of type mismatch errors between variables or literal values. And finally, semantic errors included infinite loops, array indexing errors, and division by zero errors.

AI Code Generator Usage Metrics: to measure what percentage of the code for each task was written by the learner or the AI code generator, we used the Jaccard text similarity coefficient \cite{jaccard1901etude} between each line of the final submission and the code generated by Codex, averaged over all lines. To measure what percentage of the prompt description was written by the learner themselves, or if it was simply copied from the task description, we used the same Jaccard text similarity coefficient.

\section{Results}
In this section we examine how each group performed in each of the three phases of the study. For each reported metric, we report means, standard deviations, and 95-percentile confidence intervals for each condition. An independent samples t-Test with an alpha level of 0.05 was used to determine whether there was statistical evidence that the associated population means between the two conditions were significantly different. A Bonferroni adjusted alpha level was used when performing multiple analyses at the topic level (five topics, \(\alpha=0.01\)) or error type level (three error types, \(\alpha=0.016\)). Finally, Cohen’s d is reported for effect size \cite{sullivan2012using}. We also report qualitative feedback from learners about the training phase and how learners in the Codex group felt about using the AI code generator. We report Mann-Whitney U Test to analyze statistical differences between the two groups on the Likert-scale questions, with an alpha level of 0.05 for statistical significance.

\begin{figure*}
    \centering
    \includegraphics[width=\textwidth]{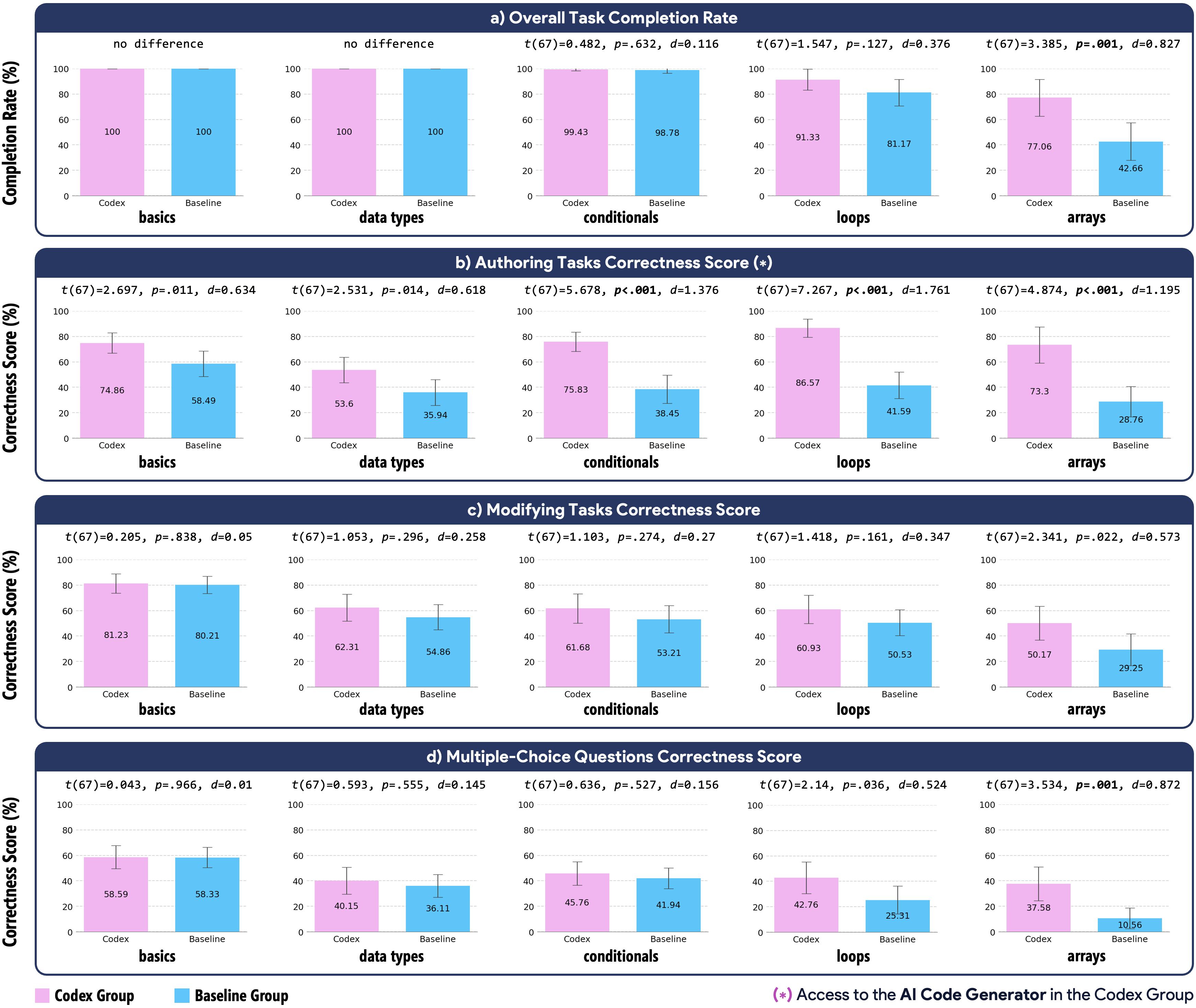}
    \caption{Correctness score of tasks during the training phase broken down by topic. Alpha levels for per-topic comparisons are adjusted using Bonferroni correction (\(\alpha=0.05/5=0.01\)).}
    \Description{The figure presents data on the performance of two groups, the Codex group and the baseline group, during the training phase. The data is presented in the form of bar graphs, with four rows and five columns of graphs. Each row represents a different aspect of performance, with the first row showing the overall task completion rate for each topic. The second row displays the correctness score of authoring tasks, while the third row displays the correctness score of modifying tasks. The fourth row shows the scores on multiple-choice questions. In general, the Codex group performed better than the baseline group, particularly on conditionals, loops, and arrays. However, both groups performed similarly on most topics, with the Codex group performing slightly better overall. The bar graphs are color-coded to differentiate between the two groups, and each bar graph includes a label indicating the topic and the performance metric. The figure provides a clear visual representation of the performance differences between the two groups, which can be used to evaluate the effectiveness of the Codex tool.}
    \label{fig:train_phase_results}
\end{figure*}

\subsection{Training Phase}
In this section we report how the two groups progressed and performed differently on three major types of tasks during the training phase: authoring, modifying and multiple-choice tasks. Figure~\ref{fig:train_phase_results} illustrates a summary of all results in this phase.

\subsubsection{Overall Completion Rates and Direct Hints}
In terms of overall progress in the training phase (Figure~\ref{fig:train_phase_results}a), learners from the Codex group finished significantly more tasks compared to the Baseline group (Codex: \textit{M}=90.9\%, \textit{SD}=16.7\%, Baseline: \textit{M}=79.0\%, \textit{SD}=18.6\%, t(67)=2.8,  \textit{p}=.006, \textit{d}=0.68). Particularly, learners from both groups fully completed the first two topics (basics and data-types) and one learner from each group did not complete conditionals. Additionally, the Codex group had more learners that fully completed the 18 tasks on loops (Codex: 27/33, Baseline: 23/36) and the 7 tasks on arrays (Codex: 23/33, Baseline: 9/36). In terms of personalized feedback during the training phase, learners in both groups received similar number of personalized feedback (Codex: \textit{M}=0.33, \textit{SD}=0.27; Baseline: \textit{M}=0.31, \textit{SD}=0.20; t(67)=0.25,  \textit{p}=.805, \textit{d}=0.06) with no difference in feedback length (Codex: \textit{M}=65.8, \textit{SD}=13.7; Baseline: \textit{M}=67.5, \textit{SD}=12.0; t(67)=0.53,  \textit{p}=.60, \textit{d}=0.13) and no significant difference in direct hints (Codex: \textit{M}=2.9, \textit{SD}=2.6; Baseline: \textit{M}=3.7, \textit{SD}=2.8; t(67)=1.15,  \textit{p}=.25, \textit{d}=0.28).

\begin{figure*}
    \centering
    \includegraphics[width=\textwidth]{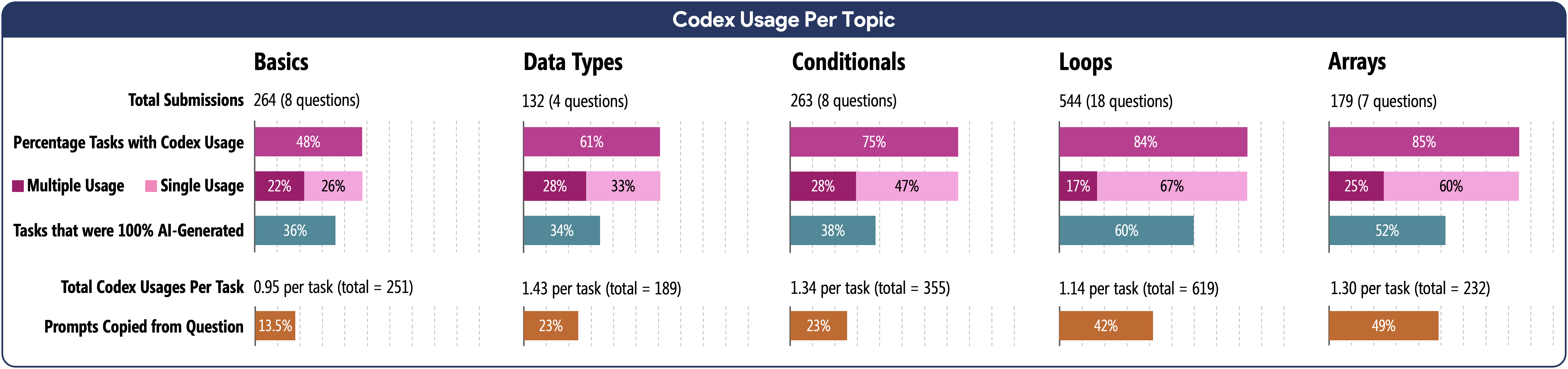}
    \caption{Statistics of tasks in which the AI Code Generator was used broken down by topic.}
    \Description{The figure displays statistics on the usage of an AI code generator for five different topics, each represented by a separate column. The graph shows that the use of the AI code generator was lower for the first topic, which covers the basics, but increased for the topics on arrays and loops. The figure also reveals that submissions that were 100\% AI-generated became more common in the later topics, and that the number of prompts copied directly from the question increased as well.}
    \label{fig:codex_usage}
\end{figure*}

\subsubsection{Authoring Tasks (Training Phase)}
On authoring tasks during the training phase, which was the only time that the Codex group had access to the AI code generator, correctness score was significantly higher for the Codex group (Codex: \textit{M}=80.1\%, \textit{SD}=14.5\%; Baseline: \textit{M}=44.4\%, \textit{SD}=26.5\%; t(67)=6.92, \textit{p}<.001; \textit{d}=1.67). Additionally, task completion time was significantly less in the Codex group (Codex: \textit{M}=210s, \textit{SD}=99s; Baseline: \textit{M}=361s, \textit{SD}=95s; t(67)=6.40,  \textit{p}<.001, \textit{d}=1.56). See Figure~\ref{fig:train_phase_results}b for more details per topic.

Furthermore, our analysis on documentation usage indicates that the Codex group accessed the documentation significantly less than the Baseline group for the authoring tasks (Codex: \textit{M}=22.1\%, \textit{SD}=21.3\%; Baseline: \textit{M}=54.3\%, \textit{SD}=26.5\%; t(67)=5.48,  \textit{p}<.001, \textit{d}=1.33).

Learners in the Codex group also encountered significantly fewer errors per task (Codex: \textit{M}=1.28, \textit{SD}=1.1; Baseline: \textit{M}=2.17, \textit{SD}=1.31; t(67)=3.03,  \textit{p}=.004, \textit{d}=0.74). Most of these errors were syntax errors that occurred significantly less for learners in the Codex group (Codex: \textit{M}=0.87, \textit{SD}=0.77; Baseline: \textit{M}=1.67, \textit{SD}=1.01; t(67)=3.65,  \textit{p}=.001, \textit{d}=0.88) followed by data-type errors in which there were no meaningful differences between the two groups (Codex: \textit{M}=0.30, \textit{SD}=0.34; Baseline: \textit{M}=0.39, \textit{SD}=0.35; t(67)=0.98,  \textit{p}=.331, \textit{d}=0.24). However, semantic errors occurred infrequently in both groups (Codex: \textit{M}=0.01, \textit{SD}=0.02; Baseline: \textit{M}=0.03, \textit{SD}=0.05; t(67)=1.84,  \textit{p}=.071, \textit{d}=0.44).

\subsubsection{AI Code Generator Usage}
In this section we report AI code generator usage and metrics in the Codex group. This data sheds some initial light on how participants incorporated the AI code generator into their workflow; additional insights are provided from qualitative feedback presented in Section~\ref{sec:qualitative_feedback}.

A summary of how learners used the AI code generator is illustrated in Figure~\ref{fig:codex_usage}. Learners used the AI code generator an average of 1.21 times per task (\textit{SD}=0.66). They did not use it at all for 26.5\% of the submitted authoring tasks (367/1382) and it was used more frequently on later topics (48\% on basics compared to 84\% on loops and arrays). Possible explanation for not using the AI code generator include their unfamiliarity with the interface or having confidence to write the code manually. Furthermore, learners used the code generator multiple times for 22.4\% of the tasks (310/1382). This could indicate that either learners needed to refine their prompt message to generate better code or that learners tried to complete the task incrementally. Follow up analysis of the submitted code indicated that in 12.9\% of the tasks (179/1382), learners broke down the task into multiple subgoals, and used the AI code generator to produce code for each requested subgoal. Future, more in-depth studies are required to reveal learners’ intentions and thought processes while using AI code generators.

To further understand how learners interacted with the AI code generator, we also analyzed each of the AI code generator prompt messages (n=1646). In particular, we found that there was a 41\% (\textit{SD}=32\%) similarity between the prompts and the task descriptions. We also found that 32\% (n=530) of the code generator prompts were exactly the same as the task description. This indicates that learners sometimes copied the task description into the code generator’s textbox.

Across all the submitted tasks in which the AI code generator was used, the final solution was on average 63\% (\textit{SD}=42\%) similar to what was generated by the AI code generator. Learners submitted the AI-generated code without any modification for 49\% of the tasks (503/1015). This particular pattern that occurred more frequently on loops (60\%) and arrays (52\%), usually happened when the learner copied the task description and asked the AI code generator to generate the solution to the whole task. However, these patterns were not consistently used across learners. In fact, 10/33 learners used this pattern for less than 4/45 of the tasks, while 14/33 learners used it for more than half of the tasks. Future qualitative analysis is required to explore ways in which learners break down tasks and write prompts for the AI code generator, or how different usage patterns impacts learning.

\begin{figure*}
    \centering
    \includegraphics[width=\textwidth]{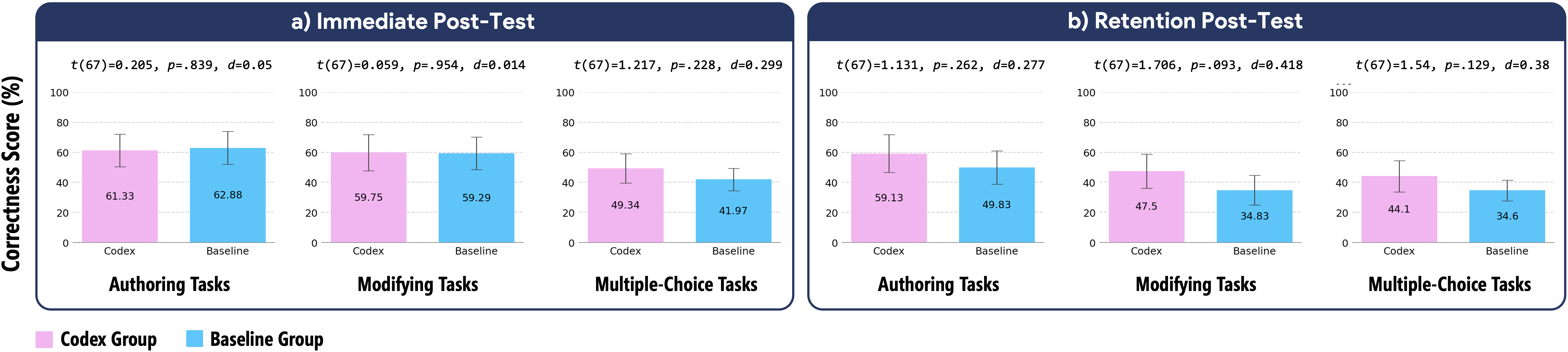}
    \caption{Correctness score of tasks during the evaluation phase.}
    \Description{The figure shows two graphs side by side, one for the immediate post-test and the other for the retention post-test. The immediate post-test graph indicates that there is not much difference in performance between the baseline group and the codex group on authoring or modifying tasks, but the codex group did slightly better on multiple-choice tasks. The retention post-test graph shows that the baseline group had a slightly greater decline in performance on all three types of tasks compared to the codex group.}
    \label{fig:eval_phase_results}
\end{figure*}

\subsubsection{Modifying Tasks (Training Phase)}
In the Modifying tasks, both groups were provided a functioning program and then were asked to modify it, with neither group having access to the AI code generator. The starter code in modifying tasks was the solution to the previous authoring tasks. We were particularly interested to see if leveraging AI code generators would be detrimental when it came to modifying existing code.

Our results (Figure~\ref{fig:train_phase_results}c) show that both groups had similar correctness scores on modifying tasks (Codex: \textit{M}=66.2\%, \textit{SD}=25.6\%; Baseline: \textit{M}=58.4\%, \textit{SD}=23.9\%; t(67)=1.28,  \textit{p}=.202, \textit{d}=0.31). In fact, learners in the Codex group on average scored 20.8\% higher on arrays in terms of correctness score (Codex: \textit{M}=50\%, \textit{SD}=36.8\%; Baseline: \textit{M}=29.2\%, \textit{SD}=36.3\%; t(39)=2.34,  \textit{p}=.022, \textit{d}=0.57) and 10.5\% higher on loops (Codex: \textit{M}=61\%, \textit{SD}=30.5\%; Baseline: \textit{M}=50.5\%, \textit{SD}=29.3\%; t(58)=1.41,  \textit{p}=.161, \textit{d}=0.34) however, neither of these results reached statistical significance. The higher completion rate on arrays in the Codex group might potentially explain this difference. But overall, this is a promising result: although learners in the Codex group used the documentation less and relied heavily on AI-generated code for the authoring tasks, they still did just as well, and in some cases better, in manual code modification tasks.

Both groups had similar documentation access rates per task (Codex: \textit{M}=13.4\%, \textit{SD}=10.5\%; Baseline: \textit{M}=17.7\%, \textit{SD}=12.3\%; t(67)=1.53,  \textit{p}=.129, \textit{d}=0.37) and similar overall errors encountered per task (Codex: \textit{M}=1.06, \textit{SD}=0.80; Baseline: \textit{M}=1.24, \textit{SD}=0.75; t(67)=0.91,  \textit{p}=.367, \textit{d}=0.22).

\subsubsection{Multiple-Choice Tasks (Training Phase)}
Both groups performed similarly on the multiple-choice questions (Figure~\ref{fig:train_phase_results}d) during the training phase (Codex: \textit{M}=48.8\%, \textit{SD}=23.5\%, Baseline: \textit{M}=45.2\%, \textit{SD}=20.2\%,  \textit{p}=.510). Looking at the correctness score on multiple-choice questions broken down by topic, we see that both groups did comparably well on topics related to basics, data-types, conditionals, and loops, however, on arrays, the Codex group performed significantly better (Codex: \textit{M}=37.6\%, \textit{SD}=36.8\%; Baseline: \textit{M}=10.5\%, \textit{SD}=23.7\%; t(39)=3.53,  \textit{p}=.001, \textit{d}=0.87).

\subsection{Evaluation Phase}
For our analysis of the evaluation phase, we excluded tasks related to topics for which the learner progressed less than 50\% during the training phase.

\subsubsection{Immediate Post-Test}
Learners in both groups performed similarly on all three types of tasks in the immediate post-test (Figure~\ref{fig:eval_phase_results}a) that was conducted a day after the training phase. Both groups performed similarly in terms of correctness score on authoring tasks (Codex: \textit{M}=61.3\%, \textit{SD}=30.1\%; Baseline: \textit{M}=62.9\%, \textit{SD}=32.0\%; t(67)=0.20,  \textit{p}=.838, \textit{d}=0.05) and modifying tasks (Codex: \textit{M}=59.7\%, \textit{SD}=33.4\%; Baseline: \textit{M}=59.3\%, \textit{SD}=31.6\%; t(67)=0.058,  \textit{p}=.953, \textit{d}=0.01). There were no meaningful differences on task completion times and error-rates on both authoring and modifying tasks between the two groups. Furthermore, on multiple-choice questions, overall, both groups scored comparably (Codex: \textit{M}=49.3\%, \textit{SD}=27.3\%; Baseline: \textit{M}=42.0\%, \textit{SD}=21.6\%; t(67)=1.21,  \textit{p}=.228, \textit{d}=0.30) while the Codex group performed significantly better on arrays (Codex: \textit{M}=32.2\%, \textit{SD}=31.8\%; Baseline: \textit{M}=13.2\%, \textit{SD}=22.8\%; t(39)=2.78,  \textit{p}=.007, \textit{d}=0.68) and on average 16\% higher on loops, but this did not reach statistical significance (Codex: \textit{M}=50.2\%, \textit{SD}=35.7\%; Baseline: \textit{M}=33.8\%, \textit{SD}=27.7\%; t(58)=2.08,  \textit{p}=.041, \textit{d}=0.51).

\subsubsection{Retention Post-Test}
The retention post-test was administered a week later to understand if AI code generators would be detrimental to learners’ ability to retain knowledge and skill. In terms of correctness score, our results (Figure~\ref{fig:eval_phase_results}b) show that learners from the Codex group on average scored 9.0\% higher on authoring tasks (Codex: \textit{M}=59.1\%, \textit{SD}=34.8\%; Baseline: \textit{M}=49.8\%, \textit{SD}=32.4\%; t(67)=1.13,  \textit{p}=.262, \textit{d}=0.28), and 12.7\% higher on modifying tasks (Codex: \textit{M}=47.5\%, \textit{SD}=31.6\%; Baseline: \textit{M}=34.8\%, \textit{SD}=29.0\%; t(67)=1.70,  \textit{p}=.092, \textit{d}=0.41), but neither of these reached statistical significance. Similar to the immediate post-test, the two groups had similar completion times on both authoring and modifying tasks. However, on error-rates, learners in the Codex group encountered significantly more errors on authoring tasks (Codex: \textit{M}=1.58, \textit{SD}=1.34; Baseline: \textit{M}=0.99, \textit{SD}=0.92; t(67)=2.08,  \textit{p}=.042, \textit{d}=0.51) that no starter code was provided. Learners in the Codex group also encountered slightly higher errors on modifying tasks, but this did not reach statistical significance (Codex: \textit{M}=0.81, \textit{SD}=0.85; Baseline: \textit{M}=0.57, \textit{SD}=0.70; t(67)=1.21,  \textit{p}=.229, \textit{d}=0.30). This increased error-rate in the Codex group could potentially be explained by the significantly higher percentage of tasks that were skipped by learners in the Baseline group compared to the Codex group (Codex: \textit{M}=14\%, \textit{SD}=35\%, Baseline: \textit{M}=33\%, \textit{SD}=47\%, t(67)=4.10,  \textit{p}<.001, \textit{d}=0.44).

For multiple-choice questions, overall both groups performed comparably well (Codex: \textit{M}=44.1\%, \textit{SD}=28.9\%; Baseline: \textit{M}=34.6\%, \textit{SD}=20.3\%; t(67)=1.54,  \textit{p}=.129, \textit{d}=0.38) while learners in the Codex group scored on average 16\% higher on loops (Codex: \textit{M}=40.6\%, \textit{SD}=34.5\%; Baseline: \textit{M}=24.7\%, \textit{SD}=26.6\%; t(58)=2.09,  \textit{p}=.04, \textit{d}=0.51) and 14\% higher on arrays (Codex: \textit{M}=30.9\%, \textit{SD}=32\%, Baseline: \textit{M}=16.7\%, \textit{SD}=24.7\%, t(39)=2.02,  \textit{p}=.047, \textit{d}=0.49), although neither of these differences reached statistical significance.

\begin{figure*}
    \centering
    \includegraphics[width=\textwidth]{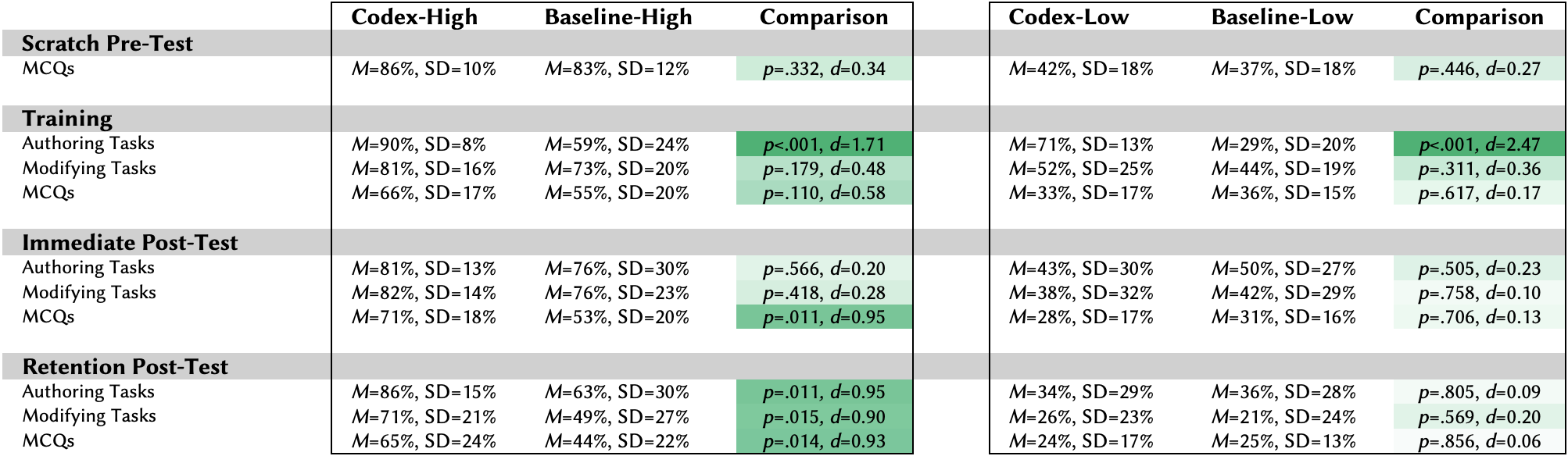}
    \caption{Correctness scores and completion rate of each quartile (based on pre-test scores and access to AI code generator). Heat-map of statistics is colored based on effect size value (greener colors indicates higher effect sizes)}
    \Description{The figure is a table that displays the correctness scores and completion rates of each quartile. The quartiles were divided based on pre-test scores and access to an AI code generator. Each row in the table is a measure compared over four graphs: Codex-high vs Baseline-high, including their means, variances, and statistics, and Codex-low vs Baseline-low statistics. The figure clearly indicates that most of the differences, especially in the retention post-test phase, are from students with higher pre-test scores. In other words, the p-values and effect sizes between the Codex-high and Baseline-high are much more significant than the differences between the Codex-low and Baseline-low, which actually performed similarly.}
    \label{fig:effect_pre_test}
\end{figure*}

\subsection{The Effect of Prior Programming Competency}
To explore how prior programming competency affects learning performance with and without access to AI code generators, we divided learners into two groups based on their pre-test scores in the introduction phase. The Codex-High (CH: n=16, \textit{M}=86.1\%, \textit{SD}=10\%) and Baseline-High (BH: n=18, \textit{M}=82.6\%, \textit{SD}=12\%) were learners that scored higher on the pre-test, and the Codex-Low (CL: n=17, \textit{M}=42.1\%, \textit{SD}=18\%) and Baseline-Low (BL: n=18, \textit{M}=37.3\%, \textit{SD}=18\%) were learners that scored lower on the pre-test.

The results of comparing each measure within these groups are provided in Figure~\ref{fig:effect_pre_test}. The comparison is showing that most of the differences between conditions (in terms of effect size) are appearing on the left side of this table, for the high performers. That is, learners in the Codex-High group performed significantly better than the Baseline-High group on multiple measures during the retention post-tests, while the Codex-Low and Baseline-Low groups had nearly similar performance levels, except on authoring tasks during the training phase. One potential explanation is that learners in the Codex-High and Baseline-High groups knew more about the fundamental concepts (e.g., conditionals and loops) and therefore, using the AI code generator provided the scaffolding needed to transition that knowledge from Scratch to Python, which helped them to perform significantly better compared to the Baseline-High group. More research is warranted to further tease out this potential effect.

\subsection{Qualitative Feedback}\label{sec:qualitative_feedback}
We inquired learners’ perception on learning, stress, and discouragement during the training phase and their eagerness about future computing education using several Likert-scale questions. We also asked learners in the Codex group several Likert-scale questions specifically about the AI code generator and a few open-ended questions on their likes and dislikes about it. A summary of the responses to the Likert-scale questions is illustrated in Figure~\ref{fig:likert_scales}.

\subsubsection{Perceptions on Learning, Stress, and Eagerness}
As shown in Figure~\ref{fig:likert_scales}a, both groups felt they learned about Python programming and its concepts during the training phase. However, on stress and discouragement, learners in the Codex group felt slightly less stressed (\textit{U}=390.5,  \textit{p}=.056). Some learners from the Codex explicitly attributed their reduced stress to using the AI Code Generator. For example, P26 reported “\textit{using the code generator helped me save time and reduced pressure}”. Additionally, learners from the Codex group felt more eager and excited to continue learning about programming after the study (\textit{U}=692,  \textit{p}=.025).

\subsubsection{Perceptions on Using AI Code Generator}
\begin{figure*}
    \centering
    \includegraphics[width=\textwidth]{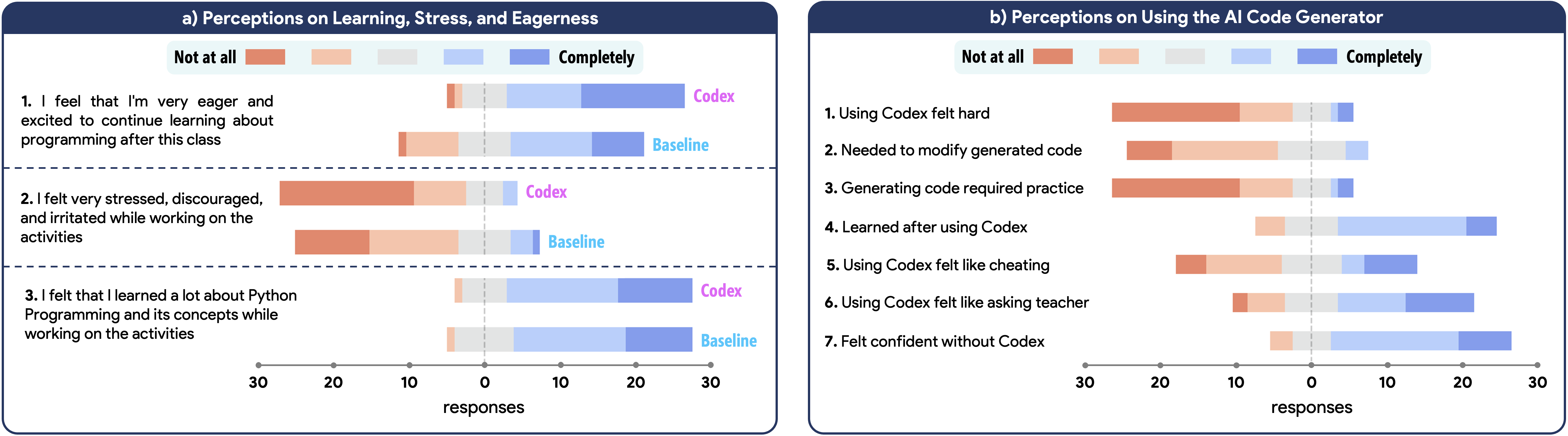}
    \caption{Aggregated responses to the Likert-scale questions.}
    \Description{The figure includes two sub-figures that display responses to Likert-scale questions. The first sub-figure is on the left, and it includes responses from both groups on questions related to eagerness to continue learning about Python, perceived stress and discouragement, and perceived learning. The values tend to be towards "not at all" for both groups on the stressed questions, while being more towards "completely" for the "I'm very eager" and "I felt that I learned a lot" Likert-questions. The sub-figure on the right includes responses to seven Likert-scales. The first question, "Using Codex felt hard," is more towards "not at all." The second question, "Needed to modify generated code," is also more towards "not at all." The third question, "Generating code required practice," is more towards "not at all." The fourth question, "Learned after using Codex," is more towards "completely." The fifth question, "Using Codex felt like cheating," has mixed results, with many towards "completely" and many towards "not at all." The sixth question, "Using Codex felt like asking teacher," is a bit more mixed but definitely more towards "completely." And the seventh question, "Felt confident without Codex," is more towards "completely."}
    \label{fig:likert_scales}
\end{figure*}

On questions about using the AI code generator (Figure~\ref{fig:likert_scales}b), learners in the Codex group generally felt that using the code generator was easy to use. For example, P4 reported “\textit{you could ask anything you want to the generator, and it would turn it into code}”. They also felt that generating the code that they needed did not require a lot of practice. P12 reported “\textit{I liked how you just had to use regular sentences like how I’m typing right now}”. Furthermore, they also did not feel the need to change many things in the AI-generated code to make it work. For instance, P19 reported “\textit{You could sometimes do an entire task without writing any code}” and P25 reported “\textit{I liked how it presented code based on the words the user had entered in the input box, generating the necessary code lines and matching it with the given variables}”. Additionally, several learners explicitly mentioned that they used the code generator whenever they were stuck. For instance, P29 reported “\textit{I liked the fact that if you were stuck, you could ask it for some code}”. A few learners also reported some difficulties. For example, P3 reported “\textit{sometimes you had to be more descriptive, and you had to say things almost to point}” and P14 reported “\textit{Sometimes I felt like it would have been easier writing the code instead of making the code generator do it}”.

Furthermore, learners mostly felt that they learned about Python programming concepts after using the code generator. For instance, P13 reported “\textit{it made it easier to learn certain applications of code such as asking it to determine if a number was even}”, and P19 reported “\textit{It would write the code for you, and you could study it if you didn’t know how}”. Several learners also mentioned that they did not like that the code generator doing the task for them without having them put any effort in the task. For example, P25 reported “\textit{What I didn’t like about it was that it was giving the direct answer to the user, instead of step-by-step hints to the user. It should give the user time to think about the problem / code before seeing the response}”. Participants also mostly felt that using the code generator was similar to asking the instructor for help. P24 reported “\textit{the code generator was a way for me to solve my own problems without having to turn to someone}”. Some learners felt that the AI-generated code needed some explanations, for example, P18 reported “\textit{I never got an explanation on why the generated code was what it was}”. Finally, learners reported mixed responses when they were asked if using the AI Coding Assistant felt like cheating.

\section{Discussion}
Our results show initial promise in using AI code generators in introductory programming settings. We reflect further on each of our research questions below.

\textbf{RQ1:} \textit{Can novices use AI code generators?} Both our quantitative and qualitative results show evidence that overall, novice learners are able to use AI code generators to successfully solve tasks. Other than 32\% of the tasks in which learners wrote the prompt description with almost no effort, they almost always actively tested the AI generated code before submission. Learners employed various methods to generate code such as breaking down the task into subgoals and using the code generator to generate code for each of the subgoals step-by-step.

\textbf{RQ2:} \textit{How does learners’ task performance differ with and without AI code generators?} Our results from code-authoring tasks during the training phase indicate that using AI code generators can significantly increase task completion, improve correctness score, reduce encountered errors, and reduce task completion time. Furthermore, our qualitative feedback shows that having access to the AI code generator reduced stress and discouragement in addition to improving their eagerness to continue programming later on.

\textbf{RQ3:} \textit{How does learners’ ability to manually modify code differ with and without AI code generators?} Our results from code-modification tasks during the training phase show that prior access to AI code generators did not reduce learners’ ability to manually modify code afterwards. In fact, we noticed improvements in code modification skills on tasks about arrays.

\textbf{RQ4:} \textit{What are the effects on learning performance and retention from using AI code generators versus not?} Our results from the immediate post-test and retention post-test show that having access to an AI code generator does not impede learning gains. In fact, we noticed that learners who performed better on the Scratch pre-test (who presumably had more prior programming knowledge) gained a significant improvement in the retention post-test after having access to the AI code generator during the training phase. Finally, it is important to note that the training experience used in our study limited the usage of AI code generators to code-authoring tasks, and learners might have potentially learned about Python programming concepts while working on the code-modification tasks.

\textbf{RQ5:} \textit{What is the relationship between existing programming competency and the effectiveness of using AI code generators in introductory programming?} Our results show that learners with higher pre-test scores benefitted more from the AI code generators compared to those who scored lower. Particularly, learners with higher prior programming competency performed significantly better on the retention test if they had access to the AI code generator during their training phase. That being said, learners who scored lower on the pre-test still performed significantly better with the AI code generator when authoring code in the training phase. This shows that AI code generators can still be an important tool to minimize frustration for even the most inexperienced users.

\subsection{How Novices Benefit from AI Coding Assistants}
The benefit of AI code generators for novice learners could be explained by the effective employment of the use-modify-create pedagogical strategy often used in introductory programming \cite{franklin2020analysis,lee2011computational}. Although learners in the Baseline group used the documentation more frequently, they had to start each task by creating a new program from scratch before getting to the modify portion of the activity, and thus encountered more errors. However, learners from the Codex group had the advantage of using code that was generated specifically for an intended behavior that they wrote. This meant that the AI Coding Assistant turned the Create task into a Use-Modify task as they were provided with functional pieces of code in response to their written description which they could then modify. Therefore, they were able to spend more time to trace, test, and learn about the structure of the generated code before moving on to modifying it in the next task. Prior work in introductory programming has shown that code-reading and tracing skills are essential to problem-solving activities including code-writing \cite{kumar2013study,lister2009further,lopez2008relationships}. This explanation also sheds light onto why learners with higher prior programming competency benefitted more from the AI code generator. These learners had the opportunity to transform their prior conceptual knowledge of variables, conditionals, loops, and lists from Scratch to Python by utilizing the use-modify-create method. Their higher prior programming knowledge meant that they probably had more code-reading and tracing skills which means that they were able to understand that code better, and then perform better during the next modification or code-creation tasks.

Furthermore, some learners in the Codex group broke down the task into multiple subgoals and then used the AI code generator to generate the Python code that would execute that particular subgoal. This way, learners had the chance to learn about the specifics of Python programming step-by-step instead of being overwhelmed with a large amount of code. Prior research has shown that subgoal-labeled instructional material reduces extraneous cognitive load imposed on novices learning programming by chunking problem-solving steps and promoting self-explanation \cite{margulieux2012subgoal}. This could potentially explain how step-by-step usage of the AI code generator could potentially reduce cognitive load and allow novices to learn faster due to more availability of their mental resources, specifically for germane cognitive load which is responsible for creating mental models and storing long-term memory \cite{kirschner2006unguided}.

\subsection{Implications for Design}
\subsubsection{Support Complete Beginners}
AI Coding Assistants could include a few features to support learners with less or no prior programming knowledge and experience. From a tool-design perspective, learners could become overwhelmed when the AI code generator produces a huge chunk of code based on their generated description. Instead, AI code generators for novices could divide the generated code into multiple segments (based on their semantic block in the abstract syntax tree) and allow the learner to spend time on each segment separately before inserting the next segment. Furthermore, these tools could accompany each of the generated code segments with line-by-line code explanations using the code-to-language capabilities of AI Coding Assistants. Additionally, documentations and worked examples could accompany the generated code, based on the keywords and programming patterns used in each code segment.

\subsubsection{Control for Over-Utilization}
One of the main concerns with the code generation capabilities of AI Coding Assistants is that they might impede learning as a result of over-utilization. Although, our results show that this might not necessarily be problematic for learners with less prior programming knowledge as they did \textit{not} perform worse compared to the Baseline group, it would not be advantageous to their learning. However, these tools could include constraints such as not allowing the learner to use the generated code in the editor before they actively engage in completing a mini-task based on the generated code. For example, this mini-task could be a Parsons problem that is automatically produced from the AI-generated code. Or a multiple-choice question that uses the same concepts in the generated code (e.g., two nested loops). Such interactions prior to using the AI-generated code could provide new possibilities for active learning and improve conceptual understandings in a personalized way.

\subsubsection{Support Writing Prompts}
When working with AI code generator, sometimes learners felt that they needed to provide a high amount of detail in the prompt description in order for the tool to generate their desired code. In these moments, they indicated that it would have been easier if they wrote the code for that task themselves. Learners could benefit from viewing a history, or a list of common phrases while they are typing the prompt to speed up the writing process (similar to automatic code completions that are used in code editors). Furthermore, novices could benefit from an interactive, dialogue-based code generation tool in which instead of writing the whole task, they would write the task through a series of smaller prompts. Similar to the implementation used in Coding Steps, such code generation bots could be context aware and allow the learner to execute commands like “\textit{if the value of variable X is greater than Y then call function F}”.

\section{Limitations and Future Work}
The results presented in this study used correctness score as one of the main variables to compare the performance of the two conditions on both coding tasks and multiple-choice questions. Particularly on coding tasks, correctness is a quantifiable variable that we defined based on our specific rubric and grading scheme. For example, on code-modification tasks during the training phase, when there was a difference of 20\% on correctness between the two groups on arrays, it means that on average learners in one group had about one fewer substantial problem per task on arrays. We could see that using more detailed rubrics such as grading at a subgoal level per submission could provide more tangible scores, however, for the sake of simplicity we decided to use the simple rubric presented in this study.

Our results show that the difference in correctness score increased at later topics such as loops and arrays. This could be due to learning effects, as learners became more familiar with using the AI code generator. However, it could also indicate that AI code generators are most useful for more complex topics. Using a fixed order of topics was needed to provide a scaffolded learning experience. Further research could be conducted to carefully study the impact that task complexity has on the benefits of AI code generators.

When discussing our results, there were several metrics in which the differences between the two groups did not reach statistical significance but there did seem to be some difference in means and confidence intervals. The lack of statistical findings for such metrics does not necessarily mean there is no effect present, it could potentially be due to the sample size of our study. Future studies with larger sample sizes, perhaps in real classroom deployments, could provide more definitive findings. As always, statistical findings should be considered with caution, and prior research has even argued for switching away from statistical significance testing in favor of reporting informative charts with effect sizes and confidence intervals and nuanced interpretations \cite{dragicevic2015hci}, which we have also tried to provide. While out of scope for our current work, we hope to perform a more thorough analysis of our qualitative data, which could also bring to light learner behaviors and explanations of our findings, to compliment are quantitative findings. This would be particularly interesting for effects related to how learners reframe and break down questions and prompts for the AI code generator. Currently, we still know very little of when novice learners used the AI code generator on each task, what their usage patterns were, and the details about how they interacted with the AI code generator. A qualitative analysis could also answer questions such as: how did learner express their intended behavior to the AI code generator?; what common themes existed in expressing their behavior?; and, what did they do when the code generator did not generate their desired code?

In terms of the target audience, our study focused on novice learners in the context of introductory programming. But there still exist many questions about the implications of AI Coding Assistants with other populations such as conversational programmers \cite{chilana2016understanding} who are motivated to improve their efficacy in technical conversations, or experienced programmers when they want to learn a new programming language or programming library. Similarly, a future study could be conducted where AI code generators are introduced in more formal learning environments such as at the high school or university level. Such a study could examine the malicious usages of AI code generators and its impact on introductory computer science education before widely integrating into curriculum. Furthermore, although our study included many non-native English speakers, but they were mostly comfortable explaining things in English. Future work could explore natural language programming with non-English speakers.

To measure learning performance, our study included evaluation post-tests that did not leave the boundaries of what learners were trained on. An interesting question would be to see how AI code generators affect transfer of learning to new topics or more complicated tasks. Furthermore, we designed the tasks used in the training phase to help novices learn about the primary syntactical and logical constructs of a text-based programming. However, another unanswered question is whether AI code generators can support algorithmic thinking and algorithm design.

Finally, our study focused only on AI code generation which is one of the main capabilities of AI Coding Assistants. However, these tools could be utilized to generate explanations of code, or they could be used when novices are stuck on a task, by automatically fixing syntax and semantic issues in code (e.g., using the \textbf{code-davinci-edit-001} model trained by OpenAI Codex) with commands like “\textit{fix the index error in my code}”.

\section{CONCLUSION}
The prevalence of Large Language Model-based AI Coding Assistants like OpenAI Codex has the potential to change how educators teach and students learn about programming. To date, the learning effects and support provided by AI code generators have not been studied in the context of introductory programming. Our results indicate that AI Coding assistants can be an asset for computer science educators and students alike: AI code generators allowed novice programmers to perform better and faster with less frustration when writing code and did not reduce their performance on manual code modification or in the subsequent absence of AI code generators. While future studies are certainly warranted, we now have encouraging evidence that integrating AI Coding Assistants into programming support tools can scaffold the learning process for novices, which could help more students to engage in programming and broaden participation in computing.

\begin{acks}
We would like to express our sincere gratitude to the after-school coding and STEM camps that helped us run our study, especially \textbf{Coder Sports} in Ottawa, and \textbf{CodeZilla}, \textbf{Hive5 Innovative Center}, and \textbf{Junior Innovators} in the Greater Toronto Area. We thank you for your invaluable assistance in recruiting participants for our research study.
\end{acks}

\balance
\bibliographystyle{ACM-Reference-Format}
\bibliography{ref-extracts}

\onecolumn
\appendix
\section{APPENDICES}
\subsection{Coding Steps Programming Tasks}
This appendix outlines the tasks involved in the code authoring and modification tasks used in the training phase of our study (using Coding Steps). The tasks are organized into five tables based on their topics: basics, data types, conditionals, loops, and arrays.

\renewcommand{\thetable}{A}
\begin{longtable}{p{0.06\linewidth} p{0.9\linewidth}}
\caption{Basics: output, variables, concatenating strings, input, arithmetic operations, and random numbers} \\
\toprule
    \textbf{Task \#} & \textbf{Task Description} \\
\toprule
\endfirsthead

\multicolumn{2}{c}%
{\tablename\ \thetable\ -- \textit{Continued from previous page}} \\
\toprule
    \textbf{Task \#} & \textbf{Task Description} \\
\toprule
\endhead

\midrule \multicolumn{2}{r}{\textit{Continued on next page}} \\
\endfoot

\bottomrule
\endlastfoot

\textbf{1A} & Write a program that will display the following message: I'm Wall-E!\\

\midrule
\textbf{1B} & Modify the given program so it displays another message after the first one: Beep Boop\\

\midrule
\textbf{2A} & Write a program that first, creates a variable called name and sets its value to Wall-E. Then, display the value of the variable\\

\midrule
\textbf{2B} & Modify the given program's variable name from name to robot\_name\\

\midrule
\textbf{3A} & Write a program that creates a variable called name and sets its value to Wall-E. Then, display the message My name is name\\

\midrule
\textbf{3B} & Modify the given program so that it displays the following message: Hi, name! Nice to meet you!\\

\midrule
\textbf{4A} & Write a program that creates a variable called name and sets its value to ro. Then, update the name variable by adding the value bot to its previous value. Finally, display the message Created: name\\

\midrule
\textbf{4B} & Modify the given program so that instead of adding bot to the name variable at once, it adds the characters b, o, and t one at a time. Print the value of the variable name after adding each of the characters. Finally, display the message Created: name\\

\midrule
\textbf{5A} & Write a program that asks the user for their name and then stores their name into a variable called name. Finally, display the message Hello, name!\\

\midrule
\textbf{5B} & Modify the following program so that it also asks the user for their family name and stores it into family\_name. Then, display the message Hello, name family\_name!\\

\midrule
\textbf{6A} & Write a program that first, creates a variable called food1 and set its value to nuts. Then, creates another variable called food2 sets it to bolts. Afterwards, create a third variable called robot\_food and sets it to the value of food1 and food2. Finally, display the message I like robot\_food. \textbf{Note:} pay attention to the space before and after the and\\

\midrule
\textbf{6B} & Modify the following program so that it includes a third food (called food3) set to screws. Then modify robot\_food to be the value of food1, food2 and food3. Finally display the message I like robot\_food.\\

\midrule
\textbf{7A} & Write a program that sets num1 to 20, and num2 to 5. Then set another variable called add to the addition of num1 and num2, sub to their subtraction, mult to their multiplication, and div to their division. Finally, display each of the add, sub, mult and div variables.\\

\midrule
\textbf{7B} & Modify the following program so that it sets a new variable called some\_num to the addition of all add, sub, mult and div. Then, in another line, update some\_num by multiplying it by 2. Finally, display the value of some\_num\\

\midrule
\textbf{8A} & Write a program that generates a random number between 1 and 10 and sets it to a variable called num. Then, display the value of num\\

\midrule
\textbf{8B} & Modify the following program so it generates a second random number between 50 and 100 and sets it to another variable named num2. Then, display the value of num2 below the value of num\\
\end{longtable}

\renewcommand{\thetable}{B}
\begin{longtable}{p{0.06\linewidth} p{0.9\linewidth}}
\caption{Data-Types: cast integer to string, and string to integer} \\
\toprule
    \textbf{Task \#} & \textbf{Task Description} \\
\toprule
\endfirsthead

\multicolumn{2}{c}%
{\tablename\ \thetable\ -- \textit{Continued from previous page}} \\
\toprule
   \textbf{Task \#} & \textbf{Task Description} \\
\toprule
\endhead

\midrule \multicolumn{2}{r}{\textit{Continued on next page}} \\
\endfoot

\bottomrule
\endlastfoot

\textbf{9A}  & Write a program that first, sets the variable num to a random number between 1 and 10. Then create another variable called message and set it to the message num is: num. Then, display the value of message\\

\midrule
\textbf{9B}  & Modify the following program by creating a second variable called num2 and setting it to the number 5. Then change message to display the following message: num is: num and num2 is: num2\\

\midrule
\textbf{10A} & Write a program that first, sets num1 to 12, and num2 to 21. Then sets a variable named message to the value num1 times num2 = (the value of num1 multiplied by num2). Finally, print message\\

\midrule
\textbf{10B} & Modify the value of message so that it displays the value of num1 times num2 like the following example. \textbf{Note:} the values of num1 and num2 can be anything and your code should work regardless of their values\\

\midrule
\textbf{11A} & Write a program that asks the user for two numbers and then displays the sum of them\\

\midrule
\textbf{11B} & Modify the following program so that after displaying the sum of num1 and num2, it would ask for another number from the user and then display the sum of all three numbers\\

\midrule
\textbf{12A} & Write a program that asks the user for four numbers and then displays the sum of them. Note that your program should only use one variable called total. The display message when asking the user to enter a new number should also include the value of total so far. At the end, it should display the value of total like this: Total: total\\

\midrule
\textbf{12B} & Modify the following program by including a variable called count that would be incremented whenever a new number is entered. The display message when asking the user to enter a new number should also include the count of numbers entered so far. Finally, it should display the total and the count like this: The sum is: total from count entries.\\
\end{longtable}

\renewcommand{\thetable}{C}
\begin{longtable}{p{0.06\linewidth} p{0.9\linewidth}}
\caption{Conditionals: if, elif, else, comparators, and logical expressions} \\
\toprule
   \textbf{Task \#} & \textbf{Task Description} \\
\toprule
\endfirsthead

\multicolumn{2}{c}%
{\tablename\ \thetable\ -- \textit{Continued from previous page}} \\
\toprule
   \textbf{Task \#} & \textbf{Task Description} \\
\toprule
\endhead

\midrule \multicolumn{2}{r}{\textit{Continued on next page}} \\
\endfoot

\bottomrule
\endlastfoot

\textbf{13A} & Write a program that first, generates a random number between 1 and 6 and assigns it to a variable called roll and then display roll. Finally, display the message rolled six only if roll is equal to six\\

\midrule
\textbf{13B} & Modify the following program so that it generates a second random number between 1 and 6 and sets it to another variable named roll2. Display both variables and finally, display the message rolled the same only if both rolls were equal\\

\midrule
\textbf{14A} & Write a program that first, generates two random numbers between 1 and 6 and check if both of the variables are greater than 3 (either 4, 5, or 6). If both are greater than 3, then first display their values and then in another line, display the message: both rolled greater than 3\\

\midrule
\textbf{14B} & Modify the following program so that it would generate a third random number called grade between 25 and 100. Then check if roll1 and roll2 are greater than 3 in addition to grade being greater than 50. If yes, display the values of each of the three variables and display the message All three above half\\

\midrule
\textbf{15A} & Write a program that asks the user to enter a number between 10 and 100. Then, check if the number is greater than 75. If it is, display the message Greater than 75; otherwise, display the message Less than 75. Note that only one of these messages should be displayed\\

\midrule
\textbf{15B} & Modify the following program so that it asks for a second number as well. Check if the first number is greater than the second. If it is, display the message First number is greater; otherwise, display the message Second number is greater. Note that only one of these messages should be displayed\\

\midrule
\textbf{16A} & Write a program that asks the user to enter a number between 0 and 100 and set it to a variable called score. Additionally, create a variable called grade and set it to an empty text. Then check if the score is less than 50, if it is, then set grade to the letter C, if it's between 50 and 75, set grade to B, otherwise, set grade to A. Then display the message Grade: grade\\

\midrule
\textbf{16B} & Modify the following program so that if the score is less than 20 set grade to F, if it's between 20 and 40 set grade to E, if it's between 40 and 60 set grade to D, if it's between 60 and 80 set grade to C, if it's between 80 and 90 set grade to B, otherwise, sets grade to A\\

\midrule
\textbf{17A} & Write a program that creates a variable called coin. Then use a random number generator to generate a number between 1 and 2. If the number is 1, set coin to heads, otherwise, set it to tails. Then display the message Coin: followed by the value of coin\\

\midrule
\textbf{17B} & Modify the program so that it would generate a random number between 1 and 7, and then display one of the days in the week based on the number generated\\

\midrule
\textbf{18A} & Write a program that gets two numbers from the user and then asks for an operator (from one of the following choices: +, -, *, and /). Then it should check which operator the user has entered, and then perform the appropriate operation. For example, if the user enters + then it should add the two numbers and display the result\\

\midrule
\textbf{18B} & Modify the program so that it would ask a third number. And then perform the appropriate operation between the three numbers for + and *. If the user enters a different operator, then display an error message: Error: Invalid operator\\

\midrule
\textbf{19A} & Ask the user to enter a number and store it in a variable called num. Check if it is even or odd. If it is odd, display the message The number num is odd otherwise display the message The number num is even. \textbf{Hint:} a number is even if the remainder of the division of the number by 2 is 0 (or in other words, it's divisible by two)\\

\midrule
\textbf{19B} & Modify the program so that it asks for another number called divisor then, checks if the entered number is divisible by the divisor. If it is, display the message The number num is divisible by divisor otherwise display the message The number num is not divisible by divisor\\

\midrule
\textbf{20A} & Set two variables called num1 and num2 to a random number between 1 and 1000 and a third variable called result to 0. Ask the user to enter one of the two options: greater, or smaller and then check which one the user has entered. (Display an error message: Invalid Option if the user didn't enter any of the two). If the user enters greater, then check if the num1 is greater than num2. If it is, set result to num1 and otherwise, set result to num2. However, if the user enters smaller, then check if the num1 is smaller than num2. If it is, set result to num1 and otherwise, set result to num2. Finally, if the user did not enter an invalid input, display the message: You entered option, and the result is result\\

\midrule
\textbf{20B} & Modify the code by adding a third option called equal that would check if the two numbers are equal or not. If they are, then display the message The numbers are equal otherwise, display three messages (each in a line): the value of num1, the value of num2, and the message The numbers are not equal\\
\end{longtable}

\renewcommand{\thetable}{D}

\begin{longtable}{p{0.06\linewidth} p{0.9\linewidth}}
\caption{Loops: for loops, range, and while loops} \\
\toprule
   \textbf{Task \#} & \textbf{Task Description} \\
\toprule
\endfirsthead

\multicolumn{2}{c}%
{\tablename\ \thetable\ -- \textit{Continued from previous page}} \\
\toprule
   \textbf{Task \#} & \textbf{Task Description} \\
\toprule
\endhead

\midrule \multicolumn{2}{r}{\textit{Continued on next page}} \\
\endfoot

\bottomrule
\endlastfoot

% Add your table content here

\textbf{21A} & Display Hello 10 times using a loop\\

\midrule
\textbf{21B} & Modify the code so that it would instead repeat the following program for 5 times: display Hello then display World!. Then finally, display Bye Bye only once afterwards\\

\midrule
\textbf{22A} & Set a variable called num to 0 and then create a loop that would add the number 5 to num for 25 times and display the value of the variable as it increases inside the loop\\

\midrule
\textbf{22B} & Modify the program so that it includes another variable that is initially set to 125. Then the loop would reduce its value by 5 for 25 times. Display the value of both variables every time their value changes in the loop\\

\midrule
\textbf{23A} & Set a variable called text to the text w. Then create a loop that would repeatedly add the letter e to the text for 5 times and displaying the text every time. After the loop, add an exclamation mark ! to the text variable and then display its value\\

\midrule
\textbf{23B} & Add another loop to the program (after the first loop) that would add the text * for 3 times to the text variable and display the text every time. Finally, after the loop, add a dot . to the text variable and display its value\\

\midrule
\textbf{24A} & Set a variable called fruits to the text I like these fruits: . Then create a loop that would repeatedly do the following things for 5 times: first, ask the user to enter a fruit name and then adding what the user entered to the fruits (separated with a space). After the loop, display the value of the fruits variable\\

\midrule
\textbf{24B} & Change the program so that when it is done with the first loop, it would then create another variable called movies and set it to I like these movies: and then use another loop that would repeat for 5 times to ask the user for their favorite movies and then add them to the variable movies one by one. After the second loop, display the value of the movies variable\\

\midrule
\textbf{25A} & Display all the numbers from 1 to 100 line by line using a loop\\

\midrule
\textbf{25B} & Modify the following program so that it would display all the numbers from 250 to 300 line by line and after each line (inside the loop), it would also display the number multiplied by 2\\

\midrule
\textbf{26A} & Ask the user to enter a number, then display all the numbers from 1 to the number entered line by line\\

\midrule
\textbf{26B} & Modify the following program so that it would ask two numbers, then display all the numbers from the first number to the second number line by line\\

\midrule
\textbf{27A} & Write a program that would ask the user to enter a number, then use a loop to calculate the total sum of all numbers from 1 to the given number (including the given number). Finally, display the total\\

\midrule
\textbf{27B} & Modify the program so that it would ask the user for two numbers and then use a loop to calculate the total sum of all numbers between the first and the second one (including both first and second numbers). (Note that the second number should always be greater than the first number). So for example, if the user enters 3 and 8, it should calculate the sum of 3, 4, 5, 6, 7, and 8. Within the loop, also display the value of the total as it increases every time. Then finally, display the total\\

\midrule
\textbf{28A} & Write a program that would calculate the sum of all even numbers between 1 to a number asked from the user (including that number). Finally, display the sum. \textbf{Hint:} a number is even when the remainder of its division by 2 is 0\\

\midrule
\textbf{28B} & Modify the loop so that it would also calculate the sum of all odd numbers between 1 to the given number at the same time. Then display two messages, first the sum of all even numbers, then the sum of all odd numbers. \textbf{Hint:} a number is even when the remainder of its division by 2 is 0, and odd when the remainder of its division by 2 is 1\\

\midrule
\textbf{29A} & Write a program that uses a while loop to repeatedly ask the user to enter a password (as a number) and check if the password is equal to 123. If it is, display the message Password is correct. If it is not, display the message Password is incorrect and ask the user to re-enter the password. The program should stop when the user enters the number 123. Finally, after the user gets the correct password, display the message Password is correct\\

\midrule
\textbf{29B} & Modify the program by changing the password from 123 to 7512 and also count the number of incorrect attempts (each time the user enters the password incorrectly). Finally display the incorrect attempts at the end of the loop\\

\midrule
\textbf{30A} & Write a program that repeatedly does the following until the user enters the number 0: ask the user for a number, and then add it up to a variable called total. If the user enters 0, display the total at the end (only once)\\

\midrule
\textbf{30B} & Modify the program so that it calculates the average of all numbers entered by the user. \textbf{Note:} the average is the sum of all numbers entered, divided by the count of numbers entered. Hint: use another variable to count the number of numbers entered by the user\\

\midrule
\textbf{31A} & Write a program that asks the user to enter a number between 1 and 100 and then display the difference of that number with the value 50. The difference between two numbers is always a positive number. \textbf{Note:} you can use the abs( ) function to calculate the positive value of any number\\

\midrule
\textbf{31B} & Modify the program so that asks the user for two numbers and then display the difference between each of the numbers. Again, note that the difference between two numbers is always a positive number\\

\midrule
\textbf{32A} & Write a guess a number game: the program will first set the variable picked\_number to a random number between 1 and 1000. Then it should repeatedly do the following until the user guesses the number (if it's equal to picked\_number): If the user guesses a number that is too high, the program should display the message The number is too high. If the user guesses a number that is too low, the program should display the message The number is too low. Finally, if the user guesses the number, the while loop should stop repeating and the program should display the message You guessed the number!\\

\midrule
\textbf{32B} & Modify the program so that it would count the number of incorrect attempts the user has made and display it at the end. Additionally, on every guess it should check if the difference between the guessed number and the picked number is less than 50. If it is, the program should display another message You are close!. \textbf{Note:} you can use the abs( ) function to calculate the positive value of any number\\

\midrule
\textbf{33A} & Write a program that would use a while loop to repeatedly ask the user to guess a number until the user enters the number 0. Inside the loop, check if the number is divisible by both 2 and 3, if it is, then display the message The number is divisible by 2 and 3 and then break out of the loop. At the end of the loop, it should simply display the message Finished loop\\

\midrule
\textbf{33B} & Modify the program by adding another if statement inside the while loop to check if the number is divisible by 5, if it is then display the message The number is divisible by 5 and then break out of the loop. The program should also include another variable called did\_break that is set to False and then set to True if one of the breaks are triggered. At the end, display the message broke out of the loop only if the variable did\_break is equal to True\\

\midrule
\textbf{34A} & Write a program that asks the user to enter a number between 1 and 100. The program should then repeatedly decrease the number by 1 until it reaches 0 and display the number each time\\

\midrule
\textbf{34B} & Modify the program so that it would first ask the user to enter a number greater than 100 and then use the loop to continuously decreases the number by 10 every time while the number is greater than 100. At this point the number should be equal or less than 100, use another loop to continuously decrease the number by 3 every time while the number is greater than zero\\

\midrule
\textbf{35A} & Write a program that generates a number between 1 and 999999. Then, it displays the number of digits in the number by repeatedly dividing the number by 10 until it reaches 0 counting the number of times it was divided. For example, 1874 ÷ 10 = 187 (first) - 187 ÷ 10 = 18 (second) - 18 ÷ 10 = 1 (third) - 1 ÷ 10 = 0 (fourth). Therefore, 1874 has four digits\\

\midrule
\textbf{35B} & Modify the program so that it displays each digit of the number from the right to the left as the number is being divided by 10. To obtain the digit, the program should use the modulus operator (\%) to obtain the remainder of the division\\

\midrule
\textbf{36A} & Write a program that includes a for loop that uses the variable i to go from 0 to 1 (including 1). Then inside the loop, have another loop that uses the variable j to go from 0 to 1 (including 1). The program should display the value of i and j every time like the provided sample\\

\midrule
\textbf{36B} & Modify the program so that the first loop would go from 0 to 10 (instead of 0 to 1) and the second loop to go from 0 to 2 instead (including 2). It should also display the message i changed to i whenever the value of i (in the outer loop) changes\\

\midrule
\textbf{37A} & Write a program that repeatedly generates a random number between 0 and 100 until the random number that it generates becomes equal to 50 (and then stop). Then display the number of attempts it took to generate the number\\

\midrule
\textbf{37B} & Modify the program so that it stops when the random number becomes equal to any of the numbers 25, 50, or 75. It should also display which number it stopped on after displaying the number of attempts after the loop\\

\midrule
\textbf{38A} & Repeatedly roll a dice for 1000 times. At the end, display the total times it rolled six\\

\midrule
\textbf{38B} & Modify the program so that it counts the number of times it rolled each of the six faces (using six variables) and then finally display the value of all six variables\\

\end{longtable}

\renewcommand{\thetable}{E}

\begin{longtable}{p{0.06\linewidth} p{0.9\linewidth}}
\caption{Arrays: initializing lists, obtaining a list’s length, appending to lists, iterating over lists} \\
\toprule
   \textbf{Task \#} & \textbf{Task Description} \\
\toprule
\endfirsthead

\multicolumn{2}{c}%
{\tablename\ \thetable\ -- \textit{Continued from previous page}} \\
\toprule
   \textbf{Task \#} & \textbf{Task Description} \\
\toprule
\endhead

\midrule \multicolumn{2}{r}{\textit{Continued on next page}} \\
\endfoot

\bottomrule
\endlastfoot

\textbf{39A} & Create a list with these values: 1, 5, 9, 13, 17, 21. Then, display the first item in the list by accessing the list using the appropriate indices. Then, display the length of the list. Hint: you should use a special function that returns the length of a list\\

\midrule
\textbf{39B} & Modify the program so that it displays the last item in the list by accessing the list using the appropriate index. You have to calculate the index of the last item of the list using the length of the list (ask yourself what the relationship between the index of the last item of a list and the length of a list is). \textbf{Note:} You must use the len function to determine the length of the list\\

\midrule
\textbf{40A} & Write a program that creates a list with the following textual values: "math", "history", "programming", and "art". Then use a while loop and an index variable to display all of the items in the list one by one\\

\midrule
\textbf{40B} & Modify the program so that it displays the items in the list in reverse order\\

\midrule
\textbf{41A} & Write a program that creates an empty list and then, inside a for loop that repeats for 10 times, ask the user to enter a number and then add it to the end of the list. At the end, display the length of the list\\

\midrule
\textbf{41B} & Rename the list to grades and create another empty list called students before the loop. Then, inside the loop, first ask the user to enter a student name (as a text/string) and then add the student name into the students list. Then, ask the user to enter a grade (as a number/integer) and then add the grade into the grades list. Finally, after the loop, display the length of both lists \\

\midrule
\textbf{42A} & Create an empty list called grades. Then, repeatedly add a random number between 50 to 100 to the list, for a random number of times (between 15 and 25). Finally, use another loop to display all the items in the grades list. \textbf{Note:} your program should use the for loop with the range function, not a while loop\\

\midrule
\textbf{42B} & Modify the code so that it defines a second list called grades2 and uses another loop to repeatedly add a random number between 1 to 10 to the grades2 list for a random number of times (between 100 to 500). In summary, your code should define two lists with random values and random lengths, then display the contents of both lists\\

\midrule
\textbf{43A} & Create a list called numbers, and then use a for loop that repeats for 5 times to repeatedly ask the user to enter a number (as an integer) and add it to the list. Then use another loop to go through the items of the numbers list and find the largest number. Finally, display the value of the largest number. (\textbf{Note:} you can NOT use the max function.)\\

\midrule
\textbf{43B} & Modify the program so that it also finds the smallest number in the list and displays it at the end. (\textbf{Note:} you can NOT use the min or max function.)\\

\midrule
\textbf{44A} & Repeatedly ask the user to enter a movie name and add it to a list called movies until the user enters stop. At the end just display how many movies the user has entered. \textbf{Note:} The list should not contain the word stop\\

\midrule
\textbf{44B} & Create another list called ratings and for each movie that is entered (that is not equal to stop), ask the user to enter a rating from 0 to 10 (as an integer) and add the number to the ratings list. At the end, display the number of movies and the number of ratings \textbf{Note:} they should have the same number of elements and stop should not be included in the movies\\

\midrule
\textbf{45A} & Create an empty list called numbers and then use a for loop that repeats for a random number of times between 50 to 75 to update the list by adding a random number between 0 to 100. Then use another loop to find the largest number in the list. Finally, display the largest number after the second loop has finished\\

\midrule
\textbf{45B} & Create another variable called smallest and use the second for loop to find the smallest number in the list in addition to the largest. At the end, display both the largest and the smallest numbers\\

\end{longtable}

\end{document}